\documentclass[12pt]{article}
\usepackage{epsfig}
\usepackage{amssymb}
\usepackage{amsthm}
\usepackage{amsmath}
\usepackage{amssymb,amsfonts}
\usepackage{graphicx}
\newcommand{\be}{\begin{eqnarray}}
\newcommand{\ee}{\end{eqnarray}}
\renewcommand{\d}{\partial}
\newcommand{\e}{\epsilon}

\newcommand{\R}{\mathbb{R}}
\newcommand{\E}{\mathbb{E}}

\newcommand{\Z}{\mathbb{Z}}

\numberwithin{equation}{section}

\begin{document}
\pagestyle{empty}
\begin{flushright}
UMN-TH-2121/02\\
{\tt hep-th/0212060}
\end{flushright}
\vspace*{5mm}

\begin{center}
{\Large\bf Anomaly Cancellation in Seven-Dimensional}\\
\vspace{0.1cm}
{\Large\bf  Supergravity with a Boundary}\\
\vspace{1.0cm}

{{\sc Tony Gherghetta}$^a$ and {\sc Alex Kehagias}$^{b,c}$}\\
\vspace{.5cm}
{\it\small {$^{a}$School of Physics and Astronomy,
University of Minnesota,\\
Minneapolis, MN 55455, USA}}\\
{\it\small {$^{b}$ Athens National Technical University,
Zografou, Greece}}\\
{\it\small {$^{c}$ Institute of Nuclear Physics, N.C.R.P.S.
Democritos,\\GR-15310, Athens, Greece}}\\
\vspace{.4cm}
\end{center}

\vspace{1cm}
\begin{abstract}
We construct a seven-dimensional brane world in a
slice of $AdS_7$, where the boundary matter content
is fixed by the cancellation of anomalies. The seven-dimensional
minimal ${\cal N}=2$ gauged supergravity is
compactified on the orbifold $S^1/\Z_2$, and the
supersymmetric bulk-boundary Lagrangian is consistently
derived for boundary vector and hypermultipets up to fermionic
bilinear terms. Anomaly cancellation then
fixes the boundary gauge coupling in terms of the seven-dimensional
Planck mass, and a topological mass parameter of the Chern-Simons
term. In addition for gauge groups containing the standard model,
anomaly cancellation restricts the gauge groups on the six-dimensional
boundaries to be only one of the exceptional groups.
There are also special values of the separation of the
two boundaries, where the boundary couplings become singular,
and lead to a possible phase transition in the boundary theory.
Furthermore, by the AdS/CFT correspondence our brane world is
dual to a six-dimensional conformal field theory, suggesting
that our bulk theory describes the strong coupling dynamics of
six-dimensional theories.

\end{abstract}

\vfill
\begin{flushleft}
\end{flushleft}
\eject
\pagestyle{empty}
\setcounter{page}{1}
\setcounter{footnote}{0}
\pagestyle{plain}

\section{Introduction}

It is an indelible fact that the particle content of the
low-energy world is anomaly free. The cancellation of anomalies
is a crucial guiding principle, especially in theories
beyond the standard model. We are accustomed to the cancellation
of anomalies in four dimensions, but as string theory has taught us the
cancellation of anomalies in higher dimensions also leads to powerful
constraints. Recently, the idea that we live on a brane
in a higher-dimensional spacetime has led to new possibilities
for physics beyond the standard model. In the brane world, where the
geometry of extra dimensions can naturally account for
hierarchies~\cite{add,rs}, one would expect that anomaly cancellation
can further lead to constraints on the matter content.

In this regard, the archetypal model is due to Horava and
Witten~\cite{HW}, where it was shown that eleven-dimensional (11D)
supergravity compactified on the orbifold $S^1/\Z_2$, uniquely
fixes the gauge group on the ten-dimensional (10D) boundaries. This
restriction arises due to the cancellation of the ten-dimensional
 anomalies. This is unlike the brane worlds constructed in five
dimensions where the boundary gauge group is not restricted by any
local anomaly, although global anomalies may impose some
constraints~\cite{Gmeiner:2002ab}. However, gravitational
anomalies also exist in six
dimensions~\cite{Alvarez-Gaume:1983ig}, and this places nontrivial
constraints on six-dimensional (6D)
theories~\cite{Salam:1985mi,bkss,dobpop,adpy}. In this paper we
will show that in seven-dimensional (7D) brane worlds the gauge group
structure and matter content on the boundaries will be similarly
restricted~\cite{Gherghetta:2002xf}. Of course, the dimensional
reduction of the Horava-Witten (HW) model automatically gives rise to
brane worlds in seven dimensions, that are anomaly free. But our analysis will
be general, and we will construct seven-dimensional brane worlds
which satisfy all the anomaly constraints and do not necessarily
arise from the dimensional reduction of the HW
model~\cite{theisen,lust}.

Our starting point will be the minimal ${\cal N}=2$ 7D gauged
supergravity. The ungauged theory is obtained from the
compactification of M-theory on $K3$ or, equivalently, from
the compactification of strongly coupled
heterotic theory on $T^3$~\cite{Witten:1995ex}. The
compactification produces twenty two vectors resulting from expanding the
eleven-dimensional three-form on the $b_2=22$ two-cycles of the
$K3$. Three of these vectors are members of the gravity multiplet,
whereas the remaining nineteen fill vector multiplets
of the ${\cal N}=2$ 7D theory. Each vector multiplet
also contains three scalars, and the 57 total scalars parametrize the
coset space $SO(19,3)/SO(19)\times SO(3)$, for which an $SO(3)\times H$
or $SO(3,1)\times H$ subgroup of $SO(19,3)$ can be gauged. The
corresponding gauged
supergravity has been constructed in Ref.~\cite{Townsend:1983kk},
and it is interesting that this theory admits a one-parameter
extension which contains a topological mass term for the
three-form. A supersymmetric gauged theory can be obtained after
introducing an appropriate potential for the scalar field
(corresponding to the $K3$ volume). The scalar potential has two
extrema, leading to either a supersymmetric or non-supersymmetric
vacuum~\cite{Mezincescu:ta}. The supersymmetric vacuum has a negative
cosmological constant implying that the vacuum in the gauged theory is not
Minkowski spacetime but rather anti-de Sitter, $AdS_7$.
The $AdS_7$ vacuum with ${\cal N}=2$ supersymmetry has been
considered in the context of the AdS/CFT correspondence
\cite{malda}, and was shown to be the supergravity dual of the 6D
${\cal N}=(0,1)$ SCFT~\cite{ferrara}.

The minimal ${\cal N}=2$ 7D gauged supergravity may be
compactified down to six dimensions on $S^1$, even in the presence
of a cosmological constant as was shown in Ref.~\cite{Giani:dw},
where the non-chiral ${\cal N}=(1,1)$ 6D theory was obtained.
However, we are interested in the chiral ${\cal N}=(0,1)$ 6D
theory because in this case vector, tensor and hypermultiplets can
couple to gravity in a way that is restricted by anomalies. In
particular, the possibility of vector multiplets on the 6D
boundaries allows one to construct theories which contain the
standard model gauge group. Thus, we need to find a way to obtain
the chiral 6D theory from the 7D gauged supergravity.

An immediate way to obtain the 6D chiral theory is to compactify
on an orbifold $S^1/\Z_2$. This is similar to what happens in the
HW model except that in our 7D scenario the vacuum is $AdS_7$, and
not Minkowski. This difference in vacua means that the boundary
branes must now have a tension. In fact, compactifying the $AdS_7$
vacuum on an orbifold is analogous to similar compactifications of
the supersymmetric Randall-Sundrum model in a slice of
$AdS_5$~\cite{gp,susyrs}. By adding suitable boundary potential
terms, which at the $AdS_7$ minimum become the brane tensions, we
will see that the vacuum of our seven-dimensional brane world
becomes a slice of $AdS_7$. Besides the localized gravity
multiplet there will also be a localized tensor and hypermultiplet
in the resulting 6D ${\cal N}=(0,1)$ chiral theory.

However, unlike the five-dimensional case, the resulting spectrum in
a slice of $AdS_7$ is anomalous because in six dimensions there are
gravitational anomalies, like in ten (and two) dimensions. In order to
cancel these anomalies we are then forced to introduce boundary fields such
as vector, tensor and hypermultiplets~\cite{Gherghetta:2002xf}. This leads
to a restriction of the possible boundary gauge groups and matter content
on the boundaries. In particular, for the case of one tensor multiplet,
we will see that for gauge groups containing the standard model,
only exceptional groups are allowed with a restriction on the number of
generations transforming in the fundamental representation.
This is one of the main results of our paper.

Furthermore, the locally supersymmetric bulk-boundary couplings are derived
for the case of boundary vector and neutral hypermultiplets. In the HW
scenario the Bianchi identity for the four-form field strength had to be
modified in order to obtain a consistent coupling between the boundary
gauge couplings and the bulk. In our scenario a similar modification for
the Bianchi identity will be needed as well. In addition the anomaly
cancellation conditions fixes a dimensionless ratio, $\eta$, as in the
HW scenario, except that $\eta$ now relates the
6D gauge coupling, the 7D gravitational constant and a topological
mass parameter of the Chern-Simon term. For the neutral
hypermultiplet in the 6D theory, whose analogous multiplet does not exist in
the 11D HW theory, we also construct the locally supersymmetric bulk-boundary
Lagrangian. In particular we will need to modify the Bianchi identity of
the two-form field strength resulting from the bulk gauge field corresponding
to the gauged R-symmetry. This modification is crucial in showing that
the scalar manifold of the boundary hypermultiplets is indeed quaternionic, as
expected for a locally supersymmetric theory. The derivation
of these bulk-boundary couplings comprises the second
main result of our paper. The case of boundary tensor multiplets is more
complicated and their couplings will be presented elsewhere.

There is also an interesting novel phenomenon in our 7D brane world
scenario, that is not found in the HW model. In the cancellation of the
mixed anomaly terms by the Green-Schwarz mechanism, the coefficient of
the boundary kinetic terms of the gauge couplings is related to the
separation between the UV and IR brane. In certain instances, there is a
critical separation for which the boundary kinetic term vanishes. This
signifies that the gauge theory on the boundary becomes infinitely
strongly coupled and suggests that a phase transition occurs which may
be related to tensionless strings. Moreover, just as there is
a dual description of the HW theory, where the strongly coupled
10D $E_8\times E_8$ heterotic string theory is described by the 11D HW
theory, we also have a similar dual picture in our model. By
the AdS/CFT correspondence~\cite{malda}, suitably modified for the addition of
boundaries~\cite{ver,Gub,adsbound}, our seven-dimensional supergravity model
is dual to a strongly coupled conformal field theory (CFT)
in six dimensions~\cite{ferrara}, where
the gauge and matter fields on the UV brane are fundamental fields added
to the CFT, while the gauge and matter fields on the IR brane are bound
states of the CFT. It remains an intriguing question to further understand
these hybrid six-dimensional conformal field theories.

The outline of this paper is as follows. In Section 2 we compactify
the 7D supergravity Lagrangian on the orbifold $S^1/\Z_2$, and identify
the surviving 6D supermultiplets. By adding boundary potential terms
we construct a 7D brane world in a slice of $AdS_7$, with localized
gravity. In general this supersymmetric brane world is anomalous, and
in Section 3 we derive the constraints needed for the cancellation of
the gauge and gravitational anomalies. In particular we will see that the
anomalies must be cancelled locally at the orbifold fixed points, as well
as globally. Explicit examples of the anomaly cancellation constraints
are given in Section 4. In Section 5
we derive the consistent locally supersymmetric bulk-boundary
Lagrangian up to bilinear fermionic terms. We consider boundary
vector multiplets and neutral boundary hypermultiplets. In the case
of boundary vector multiplets the cancellation of the anomalies fixes the
boundary gauge coupling in terms of the 7D gravitational constant, and a
topological mass parameter. For the neutral boundary hypermultiplets
we will show how the modified Bianchi identity is crucial in order
to obtain the quaternionic structure of the scalar manifold. In
Section 6 we mention the possibility of a phase transition
when the two boundaries reach a critical separation, and
comment on the dual correspondence of our model. Finally our
conclusions are presented in Section 7. Note that our conventions
and notations are summarized in Appendix A and details of the gravitational
Chern-Simons term are given in Appendix B. In Appendix
C all possible solutions satisfying the anomaly constraints with one
tensor multiplet are tabulated.

\section{The 7D supergravity Lagrangian on $S^1/\Z_2$}

The minimal ${\cal{N}}=2$ 7D  supersymmetry has an $SU(2)$
R-symmetry group and the supersymmetry multiplets are
\be
&&(A_M,{A_i}^j,\psi^i)\, , ~M,N=0,...,6, ~~i,j=1,2,
~~~~\mbox{vector multiplet}\nonumber \\
&&(g_{MN},A_{MNK},{A_{M}}^{ij},\phi,\psi_M^i,\chi^i)\, ,\,
~~~~~~~~~~~~~~~~~~~\mbox{gravity multiplet}\nonumber
\ee
The vector multiplet contains a vector $A_M$,  an
$SU(2)$ triplet of scalars ${A_i}^j$ and an $SU(2)$ pseudo-Majorana spinor
$\psi^i$, whereas the gravity multiplet contains the graviton
$g_{MN}$, an antisymmetric three-form $A_{MNK}$, an $SU(2)$
triplet of vectors ${A_M}^{ij}$, a scalar $\phi$ and the $SU(2)$
pseudo-Majorana  gravitinos $\psi_M^i$ and spinors  $\chi^i$.
The $SU(2)$ R-symmetry can be gauged and the resulting
${\cal{N}}=2$ 7D gauged supergravity with an antisymmetric two-form
$B_{MN}$ (the dual of the three-form $A_{MNK}$) has been constructed in
Refs.~\cite{Salam:1983fa,Han:1985ku}, while the one with
the three-form in Refs.~\cite{Townsend:1983kk,Mezincescu:ta,Giani:dw}.
The coupling of $n$ vector multiplets to the 7D ${\cal{N}}=2$
supergravity leads to the irreducible multiplet
$$
(g_{MN},B_{MN},A_M^I,\phi^\alpha, \psi,\chi^a, \psi_M^i)\,
,~I=1,...,n+3\, , ~a=1,...,n\, , ~\alpha=1,...,3n,
$$
where the scalars parametrize the coset $SO(n,3)/SO(n)\times
SO(3)$ as discussed in Ref.~\cite{Bergshoeff:1985mr}.

Let us now consider the pure  ${\cal N}=2$ 7D gauged supergravity
with no vector multiplets. This theory is described by the
Lagrangian~\cite{Townsend:1983kk,Mezincescu:ta,Giani:dw}
{\allowdisplaybreaks
\be
\kappa^2e^{-1}{\cal{L}}\!\!&=&
 \!\!\frac{1}{ 2}R -\frac{\sigma^{-4}}{ 48}F_{MNPQ}^2-
\frac{\sigma^{2}}{ 4}{{F_{MN}}_i}^j{{F^{MN}}_j}^i-\frac{1}{ 2}(\d_M\phi)^2
-\frac{1}{ 2}\bar{\chi}^i\Gamma^M{\cal{D}}_M\chi_i
\nonumber \\
&&\!\! -\frac{1}{ 2}\bar{\psi}_M^i \Gamma^{MNK}{\cal D}_N\psi_{Ki}
-\frac{i\sigma}{ 2\sqrt{2}}\left(\frac{1}{ 2}\bar{\psi}_K^i\Gamma^{KMNR}
\psi_{Rj}+\bar{\psi}^{Mi}\psi^{N}_j\right){{F_{MN}}_i}^j
\nonumber \\
&&\!\! -\frac{\sigma^{-2}}{ 8
\sqrt{2}}\left(\frac{1}{ 12}\bar{\psi}_K^i\Gamma^{KMNPQR}
\psi_{Ri}+\bar{\psi}^{Mi}\Gamma^{NP}\psi^{Q}_i\right)F_{MNPQ}
\nonumber \\&& \!\!
-\frac{\sigma}{
2\sqrt{10}}\bar{\chi}^i\Gamma^M\Gamma^{NK}\psi_{Mj}{{F_{NK}}_i}^j
+\frac{\sigma^{-2}}{ 24\sqrt{10}}\bar{\chi}^i\Gamma^L\Gamma^{MNPQ}
\psi_{Li}F_{MNPQ}
\nonumber \\
&& \!\!+\frac{\sigma^{-2}}{ 160\sqrt{2}}\bar{\chi}^i\Gamma^{MNPQ}\chi_iF_{MNPQ}
-\frac{3i\sigma}{
20\sqrt{2}}\bar{\chi}^i\Gamma^{MN}\chi_j{{F_{MN}}_i}^j
\nonumber \\
&& \!\! +\frac{i}{
48\sqrt{2}}F_{MNPQ}\left({{F_{KL}}_i}^j{{A_R}_j}^i-\frac{2ig}{
3}{\rm tr}(A_KA_LA_R)\right)\epsilon^{MNPQKLR}
\nonumber \\
&& \!\! +60 (m-\frac{2}{ 5}h\sigma^4)^2-10(m+\frac{8}{
5}h\sigma^4)^2+(\frac{5}{
2}m-h\sigma^4)\bar{\psi}_M^i\Gamma^{MN}\psi_{Ni}
\nonumber \\
&& \!\!+\sqrt{5}(m\!+\!\frac{8}{ 5}h\sigma^4)\bar{\psi}_M^i\Gamma^M\chi_i
\!+\!(\frac{3}{ 2}m\!+\!\frac{27}{ 5}h\sigma^4)\bar{\chi}^i\chi_i
\!+\!\frac{1}{ 2}\bar{\chi}^i\Gamma^M\Gamma^N\d_N\phi\psi_{Mi}
\nonumber \\
&& \!\! +\frac{h}{ 36} \epsilon^{KLMNPQR}F_{KLMN}A_{PQR}~, \label{GN}
\ee}
where $F_{KLMN}= 4 \partial_{[K}A_{LMN]}$ and $\kappa^2$ is the
7D Newton's constant. The local supersymmetry transformation rules are
{\allowdisplaybreaks
\be
\delta e_M^A\!\!&=&\!\!\frac{1}{ 2}\bar{\e}^i\Gamma^A\psi_{Mi}~, \label{susy1}
\\
\delta\psi_{Mi}\!\!&=&\!\!{\cal{D}}_M\e_i+\frac{\sigma^{-2}}{ 80\sqrt{2}}
\left({\Gamma_M}^{NKPQ}-\frac{8}{ 3}\delta_M^N\Gamma^{KPQ}\right)
{F_{NKPQ}}\e_i \nonumber \\
&&\!\!+\frac{i\sigma}{ 10\sqrt{2}}
\left({\Gamma_M}^{NK}-8\delta_M^N\Gamma^K\right){{F_{NK}}_i}^j\e_j
+(m-\frac{2}{ 5}h\sigma^4)\Gamma_M\e_i~,
\\
\delta A_{MNK}\!\!&=&\!\!\frac{3\sigma^{2} }{ 2\sqrt{2}}\bar{\psi}_{[M}^i
\Gamma_{NK]}\e_i+\frac{\sigma^2}{ \sqrt{10}}\bar{\chi}^i\Gamma_{MNK}\e_i~,
\\
\delta {{A_M}_i}^j\!\!&=&\!\!\frac{i\sigma^{-1} }{ \sqrt{2}}
\left(\bar{\psi}_M^j\e_i\!-\!\frac{1}{ 2}
\delta_i^j\bar{\psi}_M^k\e_k\right)\!-\!\frac{i\sigma^{-1}}{
\sqrt{10}}\left(\bar{\chi}^j\Gamma_M\e_i\!-\!\frac{1}{
2}\delta_i^j\bar{\chi}^k\Gamma_M\e_k\right)~,
\\
\delta \chi_i\!\!&=&\!\!\frac{1}{ 2}\Gamma^M\partial_M\phi\e_i-\frac{i\sigma
}{
2\sqrt{10}}\Gamma^{MN}{{F_{MN}}_i}^j\e_j+\frac{\sigma^{-2}}{
24\sqrt{10}}\Gamma^{MNPQ}F_{MNPQ}\e_i\nonumber \\
&&\!\!-\sqrt{5}(m+\frac{8}{ 5}h\sigma^4)\e_i~,
\\
\delta\phi\!\!&=&\!\!\frac{1}{ 2}\bar{\e}^i\chi_i~. \label{susyphi}
\ee}
The notation here is
\be
&&\!\!\!\!\!\!{\cal{D}}_M\chi_i=\d_M\chi_i+\frac{1}{
4}\omega_{MAB}\Gamma^{AB}\chi_i\!+\!ig {A_{Mi}}^j\chi_j\, ,
~~~m=-\frac{g\sigma^{-1}}{ 5\sqrt{2}}\, , \\
&&\!\!\!\!\!\! {{F_{MN}}_i}^j=\d_M{A_{Ni}}^j+ig {A_{Mi}}^k{A_{Nk}}^j-\!
\! M\leftrightarrow
N\, ,~~ \sigma=\exp\left(-\frac{\phi}{ \sqrt{5}}\right)~,
\label{def}
\ee
where $g$ is the $SU(2)$ coupling. The potential of the scalar $\phi$ is
\be
    V(\phi)=16h^2\sigma^8+80 h m\sigma^4-50m^2~, \label{pot}
\ee
which has (for $h/g>0$) two extrema, a non-supersymmetric local minimum
and a supersymmetric local maximum~\cite{Mezincescu:ta}. The latter is
the supersymmetric $AdS_7$ background, and so the ${\cal{N}}=2$ 7D
supergravity theory does not possess a Minkowski vacuum. Although there exists
no stable Minkowski vacuum, one can still perform a
dimensional reduction of the theory (even in the presence of a cosmological
constant). This is done by writing the 7D metric in the standard Kaluza-Klein
reduction form
\begin{eqnarray}
   ds^2_7&=&g_{MN}dx^M dx^N\nonumber \\
        &=& e^{-\xi/\sqrt{5}}g_{\mu\nu}(x^\mu)
  dx^\mu dx^\nu+e^{4\xi/\sqrt{5}}\left(dx^7+A_\mu dx^\mu\right)^2~,
\label{KKmetric}
\end{eqnarray}
where the theory is reduced along $x^7$.
For the theory (\ref{GN}), this has been performed in~\cite{Giani:dw},
and the resulting dimensional reduction produces the ${\cal{N}}=(1,1)$ 6D
supergravity theory. The 6D spectrum obtained after appropriate rescaling,
and redefinition of the various fields is
\be
(g_{\mu\nu}, \xi, A_\mu, A_{\mu\nu\rho}, A_{\mu\nu}, {A_{\mu i}}^j,
{A_i}^j,\phi, \psi_\mu^i, \psi^i, \chi^i)\nonumber~.
\ee
Before the redefinitions the 7D graviton $g_{MN}$ 
gives rise to a 6D graviton
$g_{\mu\nu}$, a scalar $g_{77}=\xi$, and a vector (graviphoton)
$g_{\mu7}=A_\mu$. From the three-form $A_{MNP}$ we get a 6D
three-form $A_{\mu\nu\rho}$ and a two-form
$A_{\mu\nu7}=A_{\mu\nu}$, from the $SU(2)$ vector ${A_{Mi}}^j$ we
get a 6D $SU(2)$ vector ${A_{\mu i}}^j$ and ${A_{7 i}}^j={A_i}^j$, from
the 7D gravitino $\psi_M^i$ we get a 6D gravitino $\psi_\mu^i$ and
a spinor $\psi_7^i=\psi^i$, while the 7D spinor $\chi^i$ gives rise
to a 6D one $\chi^i$. Dualizing the three-form $A_{\mu\nu\rho}$ into a
vector $B_\mu$ we have the ${\cal{N}}=(1,1)$ 6D massless spectrum
\be
&&(g_{\mu\nu},A_\mu,A_{\mu\nu},{A_{\mu i}}^j,\phi,\psi_\mu^i,
\chi^i)\, , ~~~~\mbox{gravity multiplet}\nonumber\\&&
(B_\mu,{A_i}^j, \xi,\psi^i)
\, , ~~~~~~~~~~~~~~~~~~~~~~\mbox{vector multiplet}\nonumber
\ee

It should be noted that the Poincar\'e (ungauged) theory is
obtained by dimensional reduction of 11D supergravity on a $K3$
surface. In this picture, the Chern-Simons term in (\ref{GN})
results from the corresponding term in 11D where the parameter $h$
is proportional to the $F_4$ ``flux'' through $K3$. The 11D
supergravity Lagrangian is invariant under $x^{11}\to -x^{11}$
provided that the three-form of the 11D gravity multiplet
transforms as $A_3\to -A_3$. Similarly, the 7D supergravity
Lagrangian, (\ref{GN}) is invariant under $x_7\to -x_7$ provided
we have \be \label{oddtran} A_{MNP}\to -A_{MNP}\, ,
~~~~{A_{Mi}}^j\to -{A_{Mi}}^j\, , ~~~~h\to -h\, , ~~~~m\to -m\, ,
\ee which is actually the transformations inherited from the 11D
parent theory. As the $\mathbb{Z}_2$ transformation $x_7\to -x_7$
is a symmetry of the theory, we can mod it out, by considering the
compactification on $S^1/\mathbb{Z}_2$. Thus the only fields which
survive at the orbifold fixed points are the $\mathbb{Z}_2$
singlets. It is not difficult to see that the $\Z_2$ parity
assignments consistent also with the supersymmetry transformations
in Eqs.(\ref{susy1})--(\ref{susyphi}) are \be &&g_{\mu\nu} ,
~A_{\mu\nu} , ~\phi ,~\xi ,~{A_i}^j ,~\psi_{\mu -}^i
, ~\chi_+^i , ~\psi_+^i ~~~~~~~~~~~\mbox{even parity}\nonumber \\
&& A_\mu, ~B_\mu,~{A_{\mu i}}^j,~\psi_{\mu +}^i,~\chi_-^i,
~\psi_-^i ~~~~~~~~~~~~~~~~~~~\mbox{odd parity}\nonumber \ee where
the $\pm$ indices refer to the chirality of the 6D reduced
spinors. In addition for the supersymmetry parameters we have that
$\epsilon_-$ is even whereas $\epsilon_+$ is odd. The odd-parity
fields are projected out while the even-parity fields survive the
orbifold projection and are organized in 6D ${\cal N}=(0,1)$
representations. At this point let us recall that the massless
representations of the $(0,1)$ supersymmetry in 6D, labeled by
their $SU(2)\times SU(2)$ representations are
$$
\begin{array}{lll} {\rm (i)}&\mbox{gravity}:& (1,1)+2 (\frac{1}{
2},1)+(0,1)\, ,\\ {\rm (ii)}& \mbox{tensor}:& (1,0)+2(\frac{1}{
2},0)+(0,0)\,,\\
 {\rm (iii)}& \mbox{vector}:&(\frac{1}{
2},\frac{1}{ 2})+2 (0,\frac{1}{ 2})\, ,\\
 {\rm (iv)}&\mbox{hyper}:&2(\frac{1}{ 2},0)+4(0,0).
\end{array}
$$
Consequently, the gravity multiplet contains the graviton, a
self-dual two-form field and a gravitino, the tensor multiplet
contains an anti-self dual two-form field, a scalar and a spinor
(tensorino), the vector contains a vector, and a gaugino, while
the hypermultiplet consists of four scalars and a spinor
(dilatino). It should be noted that all spinors are
symplectic-Majorana, and that the gaugino and gravitino have the
same chirality (left-handed), opposite to the tensorino and
dilatino (right-handed). Thus, the surviving fields on the
orbifold $S^1/\Z_2$ are arranged into the following 6D multiplets
\be &&(g_{\mu\nu},A_{\mu\nu}^+, \psi_\mu^i)\,,
~~~~~~~~~~\mbox{gravity}\label{6dmulti1}\\&& (A_{\mu\nu}^-,
\phi,\chi^i)\, ,~~~~~~~~~~~~~\mbox{tensor}\label{6dmulti2}\\&&
({A_i}^j,\xi,\psi^i)\, ,
~~~~~~~~~~~~~~\mbox{hypermultiplet}\label{6dmulti3} \ee where
$\psi_\mu^i=\psi_{\mu -}^i$ are left-handed symplectic
Majorana-Weyl fermions while $\chi^i=\chi_+^i$ and
$\psi^i=\psi_+^i$ are right-handed.

It is also interesting to point out that the 6D multiplets
(\ref{6dmulti1})-(\ref{6dmulti3}),
only exist due to the orbifold compactification. Instead if we had
compactified on $S^1$ then for nonzero $h$, the two-form $A_{\mu\nu}$
becomes massive by eating the Nambu-Goldstone bosons, $B_\mu$ in a
generalized Higgs mechanism~\cite{Giani:dw}. However by compactifying
on the orbifold, the $B_\mu$ fields are projected out and the two-form
$A_{\mu\nu}$ remains massless.

Clearly, the 6D spectrum (\ref{6dmulti1})-(\ref{6dmulti3}) at the
orbifold fixed points is anomalous and the only way to make sense of such a
theory is to introduce extra vector, hyper, and tensor multiplets at
the fixed points in such a way as to cancel any anomalous
contribution.

\subsection{7D Randall-Sundrum vacuum}

The compactification of the 7D solution on an orbifold results in
an ${\cal N}=(0,1)$ 6D theory with the massless spectrum
(\ref{6dmulti1})-(\ref{6dmulti3}), provided that under $x_7\to
-x_7$ Eq.(\ref{oddtran}) is satisfied. However, these relations do
not necessarily respect the supersymmetry transformations
(\ref{susy1})-(\ref{susyphi}) at the boundaries. For example,
since the parameters $h$ and $m$ are odd at the orbifold fixed
points, the variation of the kinetic energy terms will produce
$\delta$-function terms. In order to make the truncated theory on
the orbifold supersymmetric we must introduce six-branes at the
orbifold fixed points with specific boundary potentials. This is
very similar to the five-dimensional supersymmetric
Randall-Sundrum model~\cite{gp,susyrs}, where supersymmetry
requires the introduction of brane tensions.

In the Lagrangian (\ref{GN}) and supersymmetry transformation rules
 (\ref{susy1})-(\ref{susyphi}) let us make the replacement (with $y=x_7$)
\be
   h\to h\left[\epsilon(y)-\epsilon(y-\pi R)\right]~, \label{h}
\ee
and similarly for $m$, where $\epsilon(y)=1(-1)$ for $y>0 (y<0)$.
If we introduce the boundary potential term
\be
    S_{0} = \int d^6x \int dy \sqrt{-g}\, 20(m- \frac{2}{5}h\sigma^4)
        \left[\delta(y)-\delta(y-\pi R)\right]~, \label{Ss}
\ee
on the six-branes located at the orbifold fixed points $y^\ast$,
then the complete action will be supersymmetric.
The supersymmetric vacuum is the one in which
the Killing equations
\be
   \delta\psi_{M i}=\delta\chi_i =0~, \label{sxc}
\ee
are satisfied. Assuming that all bulk fields are zero except for the
scalar $\phi$ we find from Eq.(\ref{sxc}) that
\be
   \langle \sigma \rangle = \left(\frac{g}{8\sqrt{2} h}\right)^{\frac{1}{ 5}}~.
\ee
Substituting this vacuum expectation value
back into the bulk Lagrangian (\ref{GN}) and boundary action
(\ref{Ss}) we obtain the action
\be
\label{bbaction}
     S&=&S_7+S_{(0)}+S_{(\pi R)}~,\\
     S_7&=&\int d^6x \int dy \sqrt{-g} \left[ \frac{1}{ 2}M^5 R
         - \Lambda_7\right]~,\\
     S_{(y^\ast)}&=& \int d^6x \sqrt{-g_6} \left[ {\cal L}_{(y^\ast)}
        -\Lambda_{(y^\ast)}\right]~,
\ee
where $g_6$ is the induced metric on the six-brane located at $y^\ast$,
and $M=\kappa^{-2/5}$ is the 7D Planck mass.
The cosmological constants are given by
\be
\label{ftcond}
   \Lambda_7 = -15 M^5 k^2~~ ;~~~~~ \Lambda_{(0)}=-\Lambda_{(\pi R)}
          = 10 M^5 k~,
\ee
where
\be
\label{kdef}
    k= \left(\frac{h g^4}{ 16}\right)^{\frac{1}{5}}~.
\ee
The Einstein equations for the combined bulk and boundary action
(\ref{bbaction}) can be solved to obtain a seven-dimensional
Randall-Sundrum solution
\be
   ds^2 = e^{-2k|y|} dx_6^2 + dy^2~, \label{rs1}
\ee
where $0\leq y \leq \pi R$ and $k$ is the AdS curvature scale which
is given by (\ref{kdef}).
Note that supersymmetry automatically guarantees the fine-tuning conditions
(\ref{ftcond}) required to obtain the Randall-Sundrum solution.
This leads to a slice of $AdS_7$, where the 6D
gravity multiplet is localized on the UV brane at $y^\ast =0$. The
localization of the gravitino on the UV brane follows from the
fact that in the $AdS_7$ vacuum, the gravitino has a mass term
\be
    m_{3/2} = \frac{5}{ 2} k \left[\epsilon(y) - \epsilon(y-\pi R)\right]~,
\ee which leads to the zero mode wave function $\psi_\mu^{(0)}
\propto e^{-\frac{1}{ 2}k|y|}$ for the left-handed
$\psi_\mu^{(0)}$.

On the other hand the tensor and hypermultiplets are
localized on the IR brane. For the tensor multiplet the simplest way
to see the localization on the IR brane is to note that in the
$AdS_7$ vacuum, the scalar in the tensor multiplet has a bulk mass term
\be
     m_\phi^2 = -8k^2 + 8 k \left[\delta(y) - \delta(y-\pi
     R)\right]~,
\ee
which leads to a zero mode wavefunction $\phi^{(0)}\propto
e^{4 k |y|}$. Similarly one can check that the right-handed
tensorino, $\chi$ obtains a wavefunction $\chi^{(0)}\propto
e^{\frac{9}{ 2} k|y|}$, which is consistent with supersymmetry.
This follows from the fact that in the $AdS_7$ vacuum the
tensorino has a bulk mass term \be
     m_\chi = -\frac{3}{ 2} k \left[\epsilon(y) - \epsilon(y-\pi R)\right]~.
\ee
Thus by supersymmetry the tensor field in the tensor multiplet must be
localized on the IR brane.

For the hypermultiplet we notice that the scalar is identified as
the radion and from an analysis similar to that studied in 5d we find
that the radion is localized on the IR brane~\cite{Charmousis:1999rg}.
Thus by supersymmetry we expect the remainder of the component
fields in the hypermultiplet to be localized on the IR brane.

\section{Anomaly cancellation with a boundary}

So far the compactification of the 7D solution has resulted
in a theory with a tensor and a hypermultiplet coupled to gravity, which
are localized at one of the two boundaries. However, this 6D theory
containing only a gravity, hyper, and tensor multiplet
is anomalous. The only way to make a consistent $S^1/\Z_2$ compactification
of the 7D ${\cal N}=2$ theory is  to introduce matter on the boundaries
such that the complete theory, bulk plus boundary, is anomaly free.
Unlike the five-dimensional case where there is no anomaly constraint,
and arbitrary matter can be added to the boundaries~\cite{gp}, in our
seven-dimensional slice of AdS, we will see that anomaly cancellation
restricts the boundary matter content.
In particular, the anomaly must be cancelled both locally
on the boundaries of the 7D orbifold as well as globally
by a Green-Schwarz (GS) mechanism~\cite{gs}. Local cancellation is necessary
because otherwise the boundary theory would be
anomalous, while global cancellation is required since the 7D theory
when reduced to 6D gives rise to massless fields which contribute
to the anomaly.

\subsection{Local Green-Schwarz cancellation}

For a Green-Schwarz mechanism to take place, a necessary condition
is that the irreducible part of the anomaly cancels. Here we will
examine the cancellation of the irreducible part ${\rm tr} R^4$ of the
gravitational anomaly such that it cancels locally on
each boundary. There are four contributions to the gravitational
anomaly on each boundary coming from the gravity, vector, tensor,
and hypermultiplets. The total gravitational anomaly from
the bulk fields (the gravity, tensor and hypermultiplet) is
\be
    {\cal{A}}_{bulk}^{grav}=
      \frac{1}{ 5760}\left[243~{\rm tr} R^4
    -\frac{225}{ 4}({\rm tr}R^2)^2\right]~,
\ee
where ${\cal A}=(2\pi)^4 I_8$, with $I_8$ the anomaly eight-form.
This anomaly is distributed evenly~\cite{anomaly} between
the two boundaries at $x_7=0,\pi R$ as
\be
{\cal{A}}_{bulk}^{grav}&=&\frac{1}{2\cdot 5760}
\left[243~{\rm tr} R^4-\frac{225}{ 4}({\rm tr}R^2)^2\right]\delta(x_7)
\nonumber \\ &+&\frac{1}{2\cdot 5760}
\left[243~{\rm tr} R^4-\frac{225}{ 4}({\rm tr}R^2)^2\right]\delta(x_7-\pi R)~.
\label{anomdist}
\ee
To cancel the anomaly locally, an appropriate amount of matter
must be added to each boundary. If there are $N_V$ vectors, $N_H$ hypers and
$N_T$ tensors at $x_7=0$, and $\tilde{N}_V$ vectors, $\tilde{N}_H$
hypers and $\tilde{N}_T$ tensors at $x_7=\pi R$, then their anomaly
contribution is
\be
{\cal{A}}_{bound}^{grav}&=&
\frac{1}{ 5760}\left[(N_V\!-\!N_H\!-\!29 N_T)~{\rm tr} R^4+\frac{5}{ 4}
(N_V\!-\!N_H\!+\!7 N_T)({\rm tr}R^2)^2\right]\delta(x_7)\nonumber \\
&+&\frac{1}{5760}\left[(\tilde{N}_V\!-\!\tilde{N}_H\!-\!29 
\tilde{N}_T)~{\rm tr} R^4+\frac{5}{4}(\tilde{N}_V\!-\!\tilde{N}_H\!+\!7 
\tilde{N}_T)({\rm tr}R^2)^2\right]\!\delta(x_7-\pi R)~.\nonumber\\
\ee
Cancellation of the irreducible part of the anomaly in (\ref{anomdist}),
then leads to the conditions
\be
   N_H+29 N_T-N_V&=&\frac{243}{2}-15 n, \label{nn}\\
   \tilde{N}_H+29\tilde{N}_T-\tilde{N}_V&=&\frac{243}{2}+15 n~,
   \label{nnn}
\ee
where we have included a bulk contribution arising from a gravitational
Chern-Simons term, that only gives a meaningful solution for half-integral
values of $n$ (see Appendix B). This can also be thought of as the branes
having a ``magnetic'' charge~\cite{wittenD5}.
Thus we obtain for $N=N_H+29 N_T-N_V$, and $\tilde{N}=\tilde{N}_H+29
\tilde{N}_T-\tilde{N}_V$ the constraint $N +{\tilde N}=243$.
Then, the remaining reducible part of the total gravitational anomaly is
\be
{\cal{A}}^{grav}=\frac{1}{128}\left(N_T-4+\frac{n}{2}\right)({\rm tr}R^2)^2
\delta(x_7)+\frac{1}{128}\left(\tilde{N}_T-4-\frac{n}{2}\right)
({\rm tr}R^2)^2\delta(x_7-\pi R)~.\nonumber \\
\ee

Let us now consider the gauge and mixed anomalies.
The requirement of local anomaly cancellation for two six-branes located at
the fixed points $x_7=0,\pi R$ of the $\Z_2$ orbifold
means that the gauge group should be ${\cal G}_1\times {\cal G}_2$,
where each ${\cal G}_i$ group is localized on one of the two fixed points
(for simplicity we will only consider semisimple ${\cal G}_i$).
The pure six-dimensional anomaly
is formally described by an anomaly polynomial eight-form, $I_8$.
Thus, mathematically we require that the anomaly eight-form, $I_8$ should
satisfy
\be
\frac{\partial^2 I_8}{ \partial {\rm tr}F_1^2\partial {\rm tr}
F_2^2}=0~, \label{js}
\ee
where $F_1,F_2$ are the gauge field strengths of the
${\cal G}_1\times {\cal G}_2$ gauge group. On the 7D orbifold this condition
simply means that there is no matter charged under both
the gauge groups located at the fixed points
and the only interaction between the two six-branes is purely
gravitational. Consequently, the eight-form $I_8$ is written as
\be
I_8=I_8^{(1)}+I_8^{(2)}~, \label{local}
\ee
where $I_8^{(1)},I_8^{(2)}$ are the anomaly polynomial for the
boundary theories at $0,\pi R$, respectively.
Then, by Eq.(\ref{js}), $I_8^{(1)}$ and $I_8^{(2)}$,
appropriately normalized, are explicitly written as
\be
I_8^{(1)}&=&\left[\frac{1}{2}\left(1-\frac{N_T}{4}+\frac{n}{8}\right)
\left({\rm tr}R^2\right)^2+\frac{1}{ 6}{\rm tr}R^2 X_1^{(2)}-
\frac{2}{ 3}X_1^{(4)}\right]\delta(x_7)~,\nonumber \\
I_8^{(2)}&=&\left[\frac{1}{2}\left(1-\frac{\tilde{N}_T}{4}-\frac{n}{8}\right)
\left({\rm tr}R^2\right)^2+\frac{1}{ 6}{\rm tr}R^2 X_2^{(2)}-
\frac{2}{ 3}X_2^{(4)}\right]\delta(x_7\!-\!\pi R)~,\label{anom}
\ee
where $X_i^{(n)}$ are defined as \cite{Schwarz:1995zw}
\be
X_i^{(n)}={\rm Tr}F_i^n-\sum_in_i {\rm tr}_i F_i^n~, ~~~~i=1,2~.
\ee
As usual, ${\rm Tr}$ denotes the trace in the adjoint representation,
$ {\rm tr}_i$ the trace in the representation ${\rm R}_i$ of
the group ${\cal G}_i$, and $n_i$ is the number of hypermultiplets in the
representation  ${\rm R}_i$. For the GS mechanism  to work in this case, we
demand the local factorization (omitting  $\delta$-functions)
\be
\label{gs}
   I_8^{(i)}&=&I_4^{(i)}\,\tilde{I}_4^{(i)}~,
\ee
where $i=1,2$ and,
\be
    I_4^{(i)}=c_i{\rm tr}R^2+a_i {\rm tr}F_i^2\, ,
    ~~~~~\tilde{I}_4^{(i)}={\rm tr}R^2+b_i {\rm tr}F_i^2~,
\label{localfac}
\ee
and $c_1,c_2$ are $\frac{1}{2}(1-N_T/4+n/8)$, $\frac{1}{2}(1-
\tilde{N}_T/4-n/8)$, respectively.
Then, the anomalies $I_8^{(1)},I_8^{(2)}$ vanish by a local GS mechanism
at $x_7=0$ and $x_7=\pi R$, respectively.

A simple inspection of Eq.(\ref{anom}) reveals that such a factorization
may be problematic due to the $X^{(4)}$ term. Indeed,
\be
   X_i^{(4)}={\rm Tr}F_i^4-\sum_i n_i{\rm tr}_iF_i^4~,
\ee
is the pure gauge anomaly and  can be written as
\cite{Sagnotti:1992qw,Seiberg:1996qx},
\be
   X_i^{(4)}=\alpha_i {\rm tr}F_i^4+\gamma_i ({\rm tr}F_i^2)^2~.
\label{s4}
\ee
Similarly, we may write
\be
   X_i^{(2)}=\beta_i {\rm tr}F_i^2~.
\label{s2}
\ee
Thus, for each term (\ref{gs}) we have using (\ref{s4}) and (\ref{s2})
\be
I_8^{(i)}=c_i ({\rm tr}R^2)^2+\frac{\beta_i}{ 6}{\rm tr}R^2
{\rm tr}F_i^2- \frac{2 \alpha_i}{ 3}{\rm tr}F_i^4-\frac{2 }{
3} \gamma_i({\rm tr}F_i^2)^2~.
\ee
Then it is clear that the factorization (\ref{localfac})
is possible as long as
\be
\alpha_i {\rm tr}F_i^4=0~.
\label{a}
\ee
There are two solutions to the above equation: i) either there is
no fourth-order Casimir, or ii) $\alpha_i=0$. The first possibility
is satisfied for all the irreps of $E_8,E_7,E_6,F_4$,
$G_2,SU(3),SU(2),U(1)$, for the {\bf 28} of Sp(4)
and SU(8) and all the irreps of SO(2n) with highest weight
$(f_1,f_2,f_1,-f_2,0,...,0)$ in the Gel'fand-Zetlin basis~\cite{bkss}.
The case ii) is model dependent and should be solved in each case.

Now if (\ref{a}) is satisfied, then the anomaly can be
locally factorized as in (\ref{localfac}) with
\be
  a_i+b_ic_i=\frac{\beta_i}{ 6}\, ,~~~~ a_i b_i=- \frac{2 }{
3}\gamma_i~.
\label{sss}
\ee
Thus, $a_i,b_ic_i$ are the roots of the equation
\be
  x^2-\frac{\beta_i}{ 6}x - \frac{2}{ 3} \gamma_i c_i =0~.
\ee
This equation has real solutions (so that the anomaly can always be
factorized) for
\be
  \beta_i^2+96 \,c_i \gamma_i >0~.
\ee
Thus, for
\be
   \gamma_i>-\frac{\beta_i^2}{ 96 c_i}~,
\ee
the anomaly can be cancelled by a GS mechanism
(provided Eq.(\ref{a}) is satisfied).

\subsection{Global Green-Schwarz cancellation}

The locally factorized reducible part of the anomaly must also
cancel globally. This follows from the fact that the 7D orbifold
theory reduces in six dimensions to a Kaluza-Klein sum of massive
modes, which do not contribute to the anomaly, and the massless 6D
fields which do give an anomaly. This means that the reducible
gravitational and gauge anomalies should be cancelled by a global
GS mechanism. The irreducible part of the gravitational anomaly of
an ${\cal N}=(0,1)$ 6D supergravity theory with $n_V$ vector
multiplets, $n_T$ tensor multiplets, and $n_H$ hypermultiplets is
cancelled when~\cite{Salam:1985mi}
\be
\label{irredcond}
  n_V-n_H-29 n_T +273 =0~.
\ee
In the case of $n_T=1$, global cancellation of the anomalies leads
to the global GS condition
\be
   I_8=\left({\rm tr}R^2+ \sum_i u_i {\rm tr}F_i^2\right)
   \left({\rm tr}R^2+\sum_i v_i {\rm tr}F_i^2\right)~,
\label{global}
\ee
where $u_i,v_i$ are constants.
We will see that the combination of satisfying (\ref{gs}),
(\ref{localfac}), and (\ref{global}) leads to very stringent
possibilities for the boundary matter.

For the case of $n_T>1$ one does not require the factorization
(\ref{global}) because the presence of additional tensor fields
allows the reducible part of the anomaly to cancel via a
generalized GS mechanism~\cite{Sagnotti:1992qw}.

\section{Gauge group analysis}

As we have seen previously, the dimensional reduction of the bulk
gravity multiplet gives rise to the gravity multiplet, one tensor
multiplet, and one hypermultiplet of the ${\cal N}=(0,1)$ 6D
theory. However, this theory by itself is anomalous since
Eq.(\ref{irredcond}) is not satisfied. We are then forced to
introduce boundary fields such that in the resulting 6D theory the
anomaly can be cancelled by a GS mechanism. The fields which can
be introduced on the 6D boundaries are vector, tensor and
hypermultiplets and we will  discuss next the various
possibilities for the boundary theory.

\subsection{$n_T=1$}
In the case where there is only one tensor multiplet in the
6D theory, arising from the dimensional reduction of the bulk theory
(so that $N_T=\tilde{N}_T=0$), we are led to the constraint
\be
\label{nT1cond}
     n_H=n_V+244~.
\ee
This means that we can add vector multiplets and hypermultiplets on
the boundaries. As discussed earlier we will assume that on each boundary
there is a gauge group ${\cal G}_i$.
For the hypermultiplets charged under the gauge group ${\cal G}_i$,
we will assume that under ${\cal G}_1\times {\cal G}_2$ the total number of
hypermultiplets consist of the following representations
\be
\label{hyprep}
     n_1 (d_{F_1},1) + n_2 (1, d_{F_2}) + (n_S+1) (1,1)~,
\ee
where $d_{F_i}$ is the dimension of the fundamental representation
of the group ${\cal G}_i$, and $n_{1,2},n_S$ are constants
representing the number of each representation. Note that we have
automatically included the extra singlet hypermultiplet (or radion multiplet)
arising from the dimensionally reduced bulk theory.
Thus, assuming that the constraint (\ref{nT1cond}) is satisfied,
and simultaneously solving (\ref{localfac}) and (\ref{global})
we find the following solutions:

\begin{table}[!h]\centering
\begin{tabular}{|c|c|c|}\hline
${\cal G}_1\times {\cal G}_2$ & $n_1+n_2$ & $n_S$ \\ \hline\hline
$G_2\times G_2$ & 20 & 131\\ \hline
$F_4\times F_4$ & 10 & 87\\ \hline
$E_6\times E_6$ & 12 & 75\\ \hline
$E_7\times E_7$ & 8 & 61\\ \hline
\end{tabular}
\end{table}

We see from the table that the distribution of the 6D anomaly
on the two boundaries constrains the number of generations of
fundamental matter that can be added. In particular consider the
$E_6\times E_6$ solution. In this case we have
\begin{eqnarray}
   I_8&=&\left[{\rm tr}R^2-\frac{1}{3}({\rm tr}F_1^2 +
     {\rm tr}F_2^2)\right]
   \left[{\rm tr}R^2+(1-\frac{n_1}{6})({\rm tr}F_1^2
     -{\rm tr}F_2^2)\right]\nonumber \\
    &=&\left(c_1{\rm tr}R^2+a_1 {\rm tr}F_1^2\right)
       \left({\rm tr}R^2+b_1 {\rm tr}F_1^2\right)\nonumber\\
    &&\qquad\qquad\qquad+\left(c_2{\rm tr}R^2
       +a_2{\rm tr}F_2^2\right)
       \left({\rm tr}R^2+b_2{\rm tr}F_2^2\right)~,
\end{eqnarray}
where $n_1+n_2=12$, $c_1=1/2+n/16$, $c_2=1/2-n/16$ and $a_i=\xi_\pm$, 
$b_i c_i=\xi_\mp$ with
\be
  \xi_\pm =\frac{1}{12}\left(4-n_i\pm
       \sqrt{(4-n_i)^2+8 c_i(6-n_i)}\right)~.
\ee
In particular if one boundary contains 3 generations of the
fundamental ${\bf 27}$ then the other boundary must have 9
generations. There are also $(n_1,n_2)$ solutions $(2,7)$ and
$(5,10)$, and similar exceptions exist for the other gauge
groups. It is also possible to have two different gauge groups
distributed between the fixed points. The complete solutions
for exceptional groups can be found in Appendix C.
It should be noted that in general,
these solutions are not obtained from compactifications of
the weakly coupled heterotic $E_8\times E_8$ theory~\cite{gsw} or
from compactifications of the HW theory because the latter involves
matter charged under both local gauge groups~\cite{theisen,lust}.

The simultaneous constraint of satisfying (\ref{localfac}) and
(\ref{global}) has restricted the possible gauge group structure
on the boundaries. In particular notice that it is not possible
to have $SU(n),(n>3)$, $SO(n)$, and $Sp(n)$ on the boundaries because
in order to cancel the fourth order Casimir one needs
specific numbers of fundamentals which are incompatible with
(\ref{global}). For the $SO(n)$ groups ($n>6$) one can also try to
add matter in the spinorial representations. However, in this
case the fourth order Casimir cannot be cancelled as required
by (\ref{a}).

Finally note that the six-dimensional theory may still be ill
defined due to non-perturbative anomalies~\cite{Witten:fp,
Elitzur:1984kr,bkss}.  Global anomalies exist as long
as  $\pi_6({\cal G})$ is non-trivial. In our case
only the gauge group $G_2$ may be plagued by
global anomalies since $\pi_6(G_2)=\Z_3$. In particular, with $n_F$
fundamentals of $G_2$, the condition for the absence of global anomalies
is $n_F=1~~{\mbox mod}~~ 3$ \cite{Bershadsky:1997sb}. Thus, for the
examples tabulated in Appendix C containing the gauge group $G_2$,
the absence of non-perturbative anomalies will further
restrict the values of $(n_1,n_2)$.

\subsection{$n_T>1$}

More generally we can consider the addition of extra tensor multiplets,
on the boundaries in addition to the one arising from dimensional reduction.
This means that we must now satisfy the irreducible anomaly constraint
(\ref{irredcond}). Once this is done the remaining part of the anomaly
can only be cancelled by invoking the generalized GS mechanism where the
extra tensor multiplets are used to cancel part of the
anomaly~\cite{Sagnotti:1992qw}.

This can best be illustrated by an example.
Consider the product gauge group $SO(n_1)\times SO(n_2)$, where the
hypermultiplets are in the representation
\be
     (n_1-8) (d_{F_1},1) + (n_2-8) (1, d_{F_2}) + (n_S+1) (1,1)~.
\ee
Then by adding 3 tensor multiplets on each six-brane, the anomaly
polynomial can be factorized in the form
\begin{eqnarray}
   I_8&=&\frac{1}{4}\left\{\left[{\rm tr}R^2+2({\rm tr}F_1^2 +
     {\rm tr}F_2^2)\right]^2 -2\left(2{\rm tr}F_1^2+2{\rm tr}F_2^2\right)^2
   -4\left({\rm tr}F_1^2-{\rm tr}F_2^2\right)^2\right\}\nonumber \\
    &=&\left(c_1{\rm tr}R^2+a_1 {\rm tr}F_1^2\right)
       \left({\rm tr}R^2+b_1 {\rm tr}F_1^2\right)-3({\rm tr}F_1^2)^2
     \nonumber\\
    &&\qquad\qquad+\left(c_2{\rm tr}R^2+a_2{\rm tr}F_2^2\right)
       \left({\rm tr}R^2+b_2{\rm tr}F_2^2\right)
       -3({\rm tr}F_2^2)^2~,
\end{eqnarray}
where $c_1=1/8+n/16$, $c_2=1/8-n/16$ and $a_i=\xi_\pm$, $b_i c_i=\xi_\mp$ with
\be
  \xi_\pm =\frac{1}{2}\left(1\pm\sqrt{1-4 c_i}\right)~.
\ee
As we can see by allowing more tensor multiplets on the branes, we
obtain solutions which can involve gauge groups other than the
exceptional groups.
This implies that the class of anomaly
free 7D brane worlds is much bigger than those arising from using the usual
GS mechanism if one invokes the generalized GS mechanism.
These solutions that require a multiple number of tensor multiplets on the
branes are interesting because they are not derivable from ten-dimensional
heterotic string compactifications.

\newpage
\section{Constructing the bulk-boundary Lagrangian}

We have seen that a consistent 7D orbifold theory can be obtained by
adding boundary fields to cancel all the anomalies. Next we construct the
locally supersymmetric bulk-boundary Lagrangian. The basic idea in
theories with boundaries~\cite{HW} is to introduce a globally
supersymmetric theory at the boundary which is coupled to the bulk.
The combined theory is then clearly invariant
under local supersymmetry transformations in the bulk, and global
supersymmetry transformations on the boundary.
To construct a complete locally supersymmetric theory, bulk-boundary
interactions must be added (if possible). In the HW setup, the only
theory which can live on the boundary is a super Yang-Mills
theory since in ten dimensions this is the only supermultiplet of the
${\cal{N}}=1$ theory (besides the gravity multiplet, which in any
case exists due to the $S^1/\Z_2$ compactification). Moreover, in
the same framework, since there are no bulk or boundary
scalars to organize a perturbative expansion, everything is given
in terms of the 11D Newton's constant and a dimensionless number
($\eta$) which controls the boundary-bulk  coupling.

In our case, where the 7D supergravity Lagrangian is compactified
on $S^1/\Z_2$ with two boundaries the situation is more involved
because of basically two reasons. First, the 6D $(0,1)$ theory has
not only vector multiplets but also tensor and hypermultiplets.
The second reason is that there are scalars in the bulk in
addition to the scalars in the boundary theory. In particular, the
couplings of the bulk scalars to the boundary theory are not a
priori known. One way to construct these couplings is to use
supersymmetry because the final theory should have local $(0,1)$
supersymmetry on the boundary (after dimensional reduction). The
$(0,1)$ 6D pure supergravity theory was first considered in
Ref.~\cite{afr}. The coupling of an arbitrary number of tensor
multiplets to lowest order in the fermionic fields was considered
in~\cite{rom}, while the coupling of a single tensor multiplet and
an arbitrary number of hypers was studied in~\cite{ns}. Following
this work, it was shown in Ref.~\cite{Sagnotti:1992qw} how the
model of \cite{rom} can be coupled to vector multiplets by
employing gauge and gravitational anomaly arguments. Furthermore,
this was shown to be related to the supersymmetry anomaly in
Ref.~\cite{fms}. The complete $(0,1)$ supergravity coupled to vectors
and tensors has been constructed in \cite{frs}, and the inclusion
of hypermultiplets has  partially been obtained in \cite{ns1}.
More recently, the most general up to date supergravity theory
coupled to vectors, tensors and hypermultiplets has been given in
\cite{ric}.

Let us consider the most general globally supersymmetric theory on
the boundary coupled to gravity, which contains $n_V$ vectors, $n_T$
tensors, and  $n_H$ hypermultiplets. Then, the $n_T$ scalars of the tensor
multiplet parametrize the coset $SO(1,n_T)/SO(n_T)$ \cite{rom}, whereas
the $4n_H$ scalars of the hypermultiplets parametrize a quaternionic
manifold~\cite{bwit}. In the rigid (global) supersymmetric case, the scalar
manifold of the $n_T$ scalars turns out to be flat, while the scalars
in the hypermultiplets now parametrize a hyperk\"ahler manifold (see Table 1).
\begin{table}[!h]\centering
\begin{tabular}{||c|c|c||}\hline
& scalars in  $n_T$ tensors& scalars in $n_H$ hypers\\ \hline
global susy& $\E^{n_T}$ &$4n_H$-dim hyperk\"ahler\\ \hline
local susy & $SO(1,n_T)/SO(n_T)$& $4n_H$-dim  quaternionic\\ \hline
\end{tabular}
\caption{\it Scalar manifolds in the global and local case for tensors and
hypermultiplets}
\end{table}
Thus we see that the coupling of the 7D bulk scalars to the boundary
theory should be such that, in the reduced 6D theory the original hyperk\"ahler
scalar manifold of the  hypermultiplets should transform into a
quaternionic manifold, while the original flat $\E^{n_T}$ scalar
manifold for the scalars in the tensor multiplets should turn into
the coset $SO(1,n_T)/SO(n_T)$.

For the hypermultiplets, let us recall the corresponding situation
in the ${\cal N}=2$ 4D theory. There, the scalar manifold for the
hypermultiplets is hyperk\"ahler in the globally supersymmetric case
and becomes quaternionic in the local case. This is achieved
by introducing gravitinos which are coupled to the
$Sp(1)$-connection of the hyperk\"ahler manifold. Supersymmetry
requires that this connection is no longer flat (as was the case
for global supersymmetry). The hyperk\"ahler manifold is then
replaced by a quaternionic one and this is how, technically, the
quaternionic structure arises in the local ${\cal N}=2$ 4D theory.
In our case, the scalars in the hypermultiplets that are added to
the boundary also parametrize hyperk\"ahler manifolds, but
there are no boundary gravitinos which have to be added since
the gravitinos simply emerge from the dimensional reduction. Thus,
it appears that there is no obvious way to be consistent with 6D local
supersymmetry, since the latter demands that the hyperk\"ahler manifold must
be quaternionic.

However, recall that there is an $SU(2)$ bulk gauge field which couples to
the 7D gravitons. This plays the role of the $SU(2)(=Sp(1))$ connection and
as we will show, local supersymmetry in the $S^1/\Z_2$ compactification
of the 7D theory demands that the $SU(2)$ bulk gauge field is
related to the  $Sp(1)$ connection of the scalar manifold. This
boundary condition then determines the scalar manifold to be
quaternionic. On the other hand, it is easy to see that with $n_T$
tensors on the boundary, there exists a coupling which is consistent
with 6D local supersymmetry because in the reduced theory the scalar
manifold for the scalars in the tensors changes from a flat manifold to the
coset $SO(1,n_T)/SO(n_T)$.

\subsection{Bulk-boundary action}
The most general boundary theory contains vector, hyper and tensor multiplets.
The boundary action may collectively be written as
\be
   S_{boundary} =  S_0+S_{YM}+S_{H}+S_T~, \label{bdd}
\ee
where $S_0$ is given in (\ref{Ss}) and $S_{YM},~S_H,~S_T$ are
the actions for the vector, hyper, and tensor multiplets,
respectively. By demanding that both the bulk and boundary is
locally supersymmetric, we will determine the action for vectors and
neutral hypermultiplets. The case of gauged hypermultiplets can be
generalized from the neutral hypermultiplet case, and the tensor
multiplet case will be presented elsewhere.

\subsubsection{Boundary vector multiplets}

Let us start by considering the 6D globally supersymmetric action
for vector multiplets
\be
     S_{YM}^{(0)}=-\frac{1}{ \lambda^2}\int d^6 x \sqrt{-g}
     \left(\frac{1}{ 4}\sigma^{-2}F_{\mu\nu}^aF^{a\mu\nu}
     +\frac{1}{ 2}{\bar\lambda}^a\Gamma^\mu
     D_\mu\lambda^a\right)~,
\label{ym}
\ee
where $F^a_{\mu\nu}$ is the gauge field strength of the gauge fields
propagating on the six-brane, and $\sigma$ is defined in Eq.(\ref{def}).
The supersymmetry transformations are
\be
\delta A_\mu^a&=&\frac{1}{ 2}\, \sigma\,  \bar{\e}
\Gamma_\mu\lambda^a~,\label{susyvec1}\\
\delta \lambda^a&=&-\frac{1}{ 4}\, \sigma^{-1}\, \Gamma^{\mu\nu}
 F_{\mu\nu}^a\e~. \label{susyvec2}
\ee
In order to make the combined action $S_{bulk}+S_{boundary}$
locally supersymmetric, where the bulk Lagrangian is defined
by (\ref{GN}), we need to add an interaction
$\bar\psi {\cal S}_{YM}$, where ${\cal S}_{YM}$ is the
supercurrent of the vector supermultiplet. This interaction is
\be
   S^{(1)}_{YM} = -\frac{1}{4\lambda^2} \int d^6x \sqrt{g}\,\sigma^{-1}
 \bar{\psi}_\mu\Gamma^{\nu\rho}\Gamma^\mu \lambda^a
 F_{\nu\rho}^a~,
\ee
and its variation  cancels terms in
${\cal{L}}_{YM}$ of the form $D\epsilon \lambda F$ and $\epsilon \psi
\lambda D \lambda$,  whereas the uncancelled part is
\be
   \Delta^{(1)}= \frac{1}{ 16\lambda^2}\int d^6x \sqrt{g} \sigma^{-2}
     {\bar\psi}_\mu^i\Gamma^{\mu\nu\rho\sigma\tau}F_{\nu\rho}^a
     F_{\sigma\tau}^a \epsilon_i~.
\ee
Analogous to the 11D HW theory we can cancel this contribution from
the variation of the bulk term, $\bar\psi_A\Gamma^{ABCDEF}\psi_F F_{BCDE}$
by modifying the Bianchi identity as
\be
\label{bianchi}
   d F_{7\mu\nu\rho\sigma}= -3\sqrt{2}\frac{\kappa^2}{ \lambda^2}
    \delta(x_7) F_{[\mu\nu}^aF_{\rho\sigma]}^a~.
\label{BB}
\ee
There is one more term in the variation of ${\cal{L}}_{YM}$ which
remains to be cancelled. This term comes from $\delta A^a_\mu$ and
$\delta\lambda$, proportional to $\sigma$ and $\sigma^{-1}$, respectively, and it is
\be
\delta {\cal{L}}_{YM}^{(0)}=\frac{1}{ 4 \sqrt{5} \lambda^2}\int d^6x\sqrt{-g}
 \sigma^{-1} \bar{\lambda}^aF^a_{\mu\nu}
\Gamma^{\mu\nu}\Gamma^\kappa\epsilon\,\partial_\kappa\phi~.
\ee
It  can be cancelled by adding to the boundary action the term
\be
 S^{(2)}_{YM} = -\frac{1}{ 2 \sqrt{5}\lambda^2}\int d^6x\sqrt{-g}
 \sigma^{-1} \bar{\lambda}^aF^a_{\mu\nu}
\Gamma^{\mu\nu}\chi~,
\ee
where $\chi$ is the  partner of the bulk $SU(2)$ vector multiplet.
In the  variation of $ S^{(2)}_{YM}$, terms of the form
$\epsilon F^2 \chi$ are cancelled
from the variation of $\sigma^{-2}$ in  ${\cal{L}}_{YM}$.
The only uncancelled variation of  $ S^{(2)}_{YM}$ is
\be
\Delta^{(2)}=-\frac{1}{ 8 \sqrt{5}\lambda^2} \int d^6x
     \sqrt{g}\,\sigma^{-2}\bar{\epsilon}
\Gamma^{\mu\nu\rho\sigma}F_{\mu\nu}^aF_{\rho\sigma}^a\chi~.
\ee
This term should also cancel from a bulk contribution. Indeed,
in the bulk action (\ref{GN}) there exists the term
$\chi\psi_L F_{MNPQ}$, and its variation  will contain
the term $\chi D_L\epsilon F_{MNPQ}$ which after partial integration
gives
$\chi \epsilon dF$. In the usual theory without  boundaries, this
contribution vanishes due to the Bianchi identity. However,
in the presence of the boundary, there is a contribution
\be
-\frac{1}{ 24 \sqrt{10}\kappa^2}\sigma^{-2}\bar{\chi}\Gamma^{KLMNP}\epsilon
\partial_KF_{LMNP}=\frac{1}{ 8\sqrt{5}\lambda^2}\sigma^{-2}
\Gamma^{\mu\nu\rho\sigma}F_{\mu\nu}^aF_{\rho\sigma}^a\chi~,
\ee
as a result of (\ref{BB}) and it is this bulk contribution which exactly cancels
$\Delta^{(2)}$. It should be noted that there are further terms in the variation of
$S^{(1)}_{YM}$ and $S^{(2)}_{YM}$ which have to be checked. They
arise from variations $\psi_\mu$ and $\chi$ proportional to
$F_{\mu\nu\rho7}$ and ${F_{\mu 7i}}^j$, and give  terms of the
form $\epsilon F F_{\mu\nu\rho7}$ and $\epsilon F F_{\mu7}$.
In particular, the only uncancelled variation of $S^{(1)}_{YM}$
and $S^{(2)}_{YM}$ so far is
\be
\Delta^{(3)}=\frac{1}{ 48\sqrt{2}\lambda^2}\int d^6x \sqrt{-g}
\sigma^{-3} \bar{\epsilon}\left(\Gamma^{\nu\rho\sigma}
\Gamma^{\alpha\beta} +12 \Gamma^\nu\Gamma^\rho\Gamma^\alpha
g^{\sigma\beta}\right)\lambda_i^a F_{\alpha\beta}
F_{7\nu\rho\sigma}~.
\ee
There is an additional contribution from the variation of the bulk
$F_{\mu\nu\rho7}F^{\mu\nu\rho7}$ term as in the HW case. The
total variation can be cancelled by adding the term
\be
S^{(3)}_{YM}=\frac{1}{ 24\sqrt{2}\lambda^2}\int d^6 x
     \sqrt{-g}\, \sigma^{-2}
\bar{\lambda}^a\Gamma^{\mu\nu\rho}\lambda^a F_{\mu\nu\rho 7}~.
\ee
Finally, the uncancelled variation proportional to
$\epsilon F F_{\mu7}$ is
\be
\Delta^{(4)}=\frac{i}{ 4\sqrt{2}\lambda^2}\int d^6 x \sqrt{-g}\,
     \bar{\epsilon}
\Gamma^{\rho\sigma}\Gamma^\mu F^a_{\rho\sigma}F_{\mu 7}\lambda^a~,
\ee
which can be eliminated by adding the extra part
\be
S^{(4)}_{YM}=-\frac{i}{ 2\sqrt{2}\lambda^2}\int d^6x \sqrt{-g}\,\sigma
     \bar{\lambda}^{ai}\Gamma^\mu{F_{\mu 7i}}^j\lambda^a_j~.
\ee
Thus, the  bulk and boundary actions (\ref{GN}) and (\ref{bdd})
together with
\be
&& S_{YM}= -\frac{1}{ \lambda^2}\int d^6 x \sqrt{-g}
\left[\frac{1}{ 4}\sigma^{-2}F_{\mu\nu}^aF^{a\mu\nu}
     +\frac{1}{ 2}{\bar\lambda}^a\Gamma^\mu
D_\mu\lambda^a\right.\nonumber \\
&& \left.\phantom{xxXx}+\frac{1}{ 4} \sigma^{-1}
    \bar{\psi}_\mu\Gamma^{\nu\rho}\Gamma^\mu \lambda^a F_{\nu\rho}^a
+\frac{1}{ 2\sqrt{5}} \sigma^{-1} \bar{\lambda}^aF^a_{\mu\nu}
\Gamma^{\mu\nu}\chi \right.\nonumber \\
&& \left.\phantom{xxXx}
-\frac{1}{ 24 \sqrt{2}}\, \sigma^{-2}
\bar{\lambda}^a\Gamma^{\mu\nu\rho}\lambda^a F_{\mu\nu\rho 7}+
\frac{i}{ 2\sqrt{2}}\,\sigma
     \bar{\lambda}^{ai}\Gamma^\mu{F_{\mu 7i}}^j\lambda^a_j
\right]~,
\ee
are invariant under the supersymmetry transformations
(\ref{susy1})-(\ref{susyphi}) and ({\ref{susyvec1})-({\ref{susyvec2})
up to fermionic bilinear terms.

Having obtained the locally supersymmetric action we can now obtain
a relation between the boundary gauge coupling, the 7D Planck mass,
and the mass parameter of the Chern-Simons term, by explicitly
cancelling all the anomalies via the GS mechanism. First we must
solve the modified Bianchi identity (\ref{bianchi}), and
as in \cite{HW} we introduce
\be
\omega_{\mu\nu\rho}={\rm tr}\left(A_\mu F_{\nu\rho}-\frac{1}{ 3} A_\mu\
[A_\nu,A_\rho]+\mbox{cyclic perm.}\right)~,
\ee
which satisfies
\be
\partial_\lambda\omega_{\mu\nu\rho}+\mbox{cyclic perm.}=6
{\rm tr }F_{[\lambda\mu}F_{\nu\rho]}~.
\ee
Then, (\ref{bianchi}) is satisfied if $F_{7\mu\nu\rho}$ is
defined as
\be
F_{7\mu\nu\rho}=4\partial_{[7}A_{\mu\nu\rho]}+\frac{\kappa^2}{
\sqrt{2}\lambda^2}\delta(x_7)\omega_{\mu\nu\rho}~.
\ee
Since under infinitesimal gauge transformations,
$\delta A^a_\mu=-D_\mu \varepsilon^a$, $\omega$ transforms as
\be
\delta\omega_{\mu\nu\rho}=3\partial_{[\mu}{\rm tr}
(\varepsilon F_{\nu\rho]})~,
\ee
gauge invariance of $F_{7\mu\nu\rho}$ is achieved if
$A_{7\mu\nu}$ transforms as
\be
\delta A_{7\mu\nu}=\frac{\kappa^2}{ \sqrt{2}\lambda^2}\delta(x_7)
{\rm tr}(\varepsilon F_{\mu\nu})~. \label{scc}
\ee
Solving the the Bianchi identity in the ``upstairs'' approach, or
specifying  the boundary behaviour of $F_4$ in the ``downstairs''
version~\footnote{In the downstairs approach, the theory is defined
on $M^6\times S^1/\Z_2$, whereas in the upstairs version it is defined
on $M^6\times S^1$ with $\Z_2$-symmetric fields. The volume integrals
in the former case is half the integrals in the latter one~\cite{HW},
which means that $\kappa^2$ is replaced by $\kappa^2/2$ in (\ref{GN}).}
we find,  as in the HW case,  that $F_4$  has a jump at $x_7=0$ given by
\be
F_{\mu\nu\kappa\lambda}=-\frac{3}{
\sqrt{2}}\frac{\kappa^2}{ \lambda^2}\epsilon(x_7)
{\rm tr}F_{[\mu\nu}F_{\kappa\lambda]}~,
\ee
and similarly at $x_7=\pi R$.
At the classical level, the transformation of the three-form under
gauge transformations (\ref{scc}) makes the Chern-Simons term in the
7D Lagrangian (\ref{GN}) not gauge invariant, but
we will see that at the quantum level an anomalous fermion contribution
will cancel the non-gauge invariant term and restore gauge invariance, much
like in the 11D HW theory. We will next consider the anomaly cancellation
for both the gravitational and mixed anomalies.  In order to do this, it is
more convenient to use the downstairs approach in form notation.

The Chern-Simons term in the ``downstairs'' approach on the
seven-dimensional manifold $M^7$ is
\be
S_{CS}=\frac{2}{ \kappa^2}\int_{M^7} h~ F_4\wedge A_3~,
\label{CSS}
\ee where
$A_3$ is the three-form gauge field of 7D supergravity and
$F_4=dA_3$. On each component of the boundary, $\d M^7$, we will
have
\be
F_4|_{\d M^7}=\frac{\kappa^2}{ \sqrt{2} \lambda^2} Q_4~,
\ee
where the four-form $Q_4$ is defined as
\be
Q_4=\xi_{CS} \, {\rm tr} R^2-{\rm tr}F^2~,
\ee
and $\xi_{CS}$ is a numerical constant.
We may now define $Q_3=\xi_{CS} \,  \omega_{3L}-\omega_{3Y}$
where as usual $\omega_{3Y,L}$ are the Yang-Mills and Lorentz
Chern-Simons terms
\begin{eqnarray}
 \omega_{3Y}&=&{\rm tr} \left(A F-\frac{1}{ 3}A^3 \right)~,\\
 \omega_{3L}&=&{\rm tr} \left(\omega
 R-\frac{1}{
  3}\omega^3 \right)~.
\end{eqnarray}
Then, we have the descent equations
\be
Q_4=dQ_3\, , ~~~\delta Q_3=d Q_2^1~,
\ee
for $\delta$ gauge and Lorentz transformations which follows from
\be
&&d\omega_{3L}={\rm tr }R^2\, , ~~~ \delta\omega_{3L}=d
\omega_{2L}^1~,\nonumber \\&&
d\omega_{3Y}={\rm tr }F^2\, , ~~~ \delta\omega_{3Y}=
d\omega_{2Y}^1~.\label{dd}
\ee
In the following we will not need the explicit forms of
$\omega_3,\omega_2^1$. Then, following~\cite{deAlwis:1996hr,con,har}
we have
\be
  A_3|_{\d M^7}=\frac{\kappa^2}{ \sqrt{2} \lambda^2}Q_3~,
\ee
so that
\be
\delta A_3|_{\d M^7}= \frac{\kappa^2}{ \sqrt{2} \lambda^2}d Q_2^1~.
\ee
This variation is extendable to the bulk by writing
\be
\delta A_3|_{\d M^7}=d\Lambda \, , ~~~ \Lambda|_{\d M^7}=
\frac{\kappa^2}{ \sqrt{2} \lambda^2}Q_2^1~.
\ee
Then, the anomalous variation of the bulk action is
\be
\delta S_{CS}=-\frac{\kappa^2 h}{  \lambda^4}\int_{M^6}
Q_2^1\wedge Q_4~, 
\ee 
where $M^6$ is the boundary at $x_7=0$, and it should be compensated by 
the anomaly of the boundary theory. Thus, the bulk anomaly eight-form for 
the Chern-Simons term $S_{CS}$ (\ref{CSS}) is
\be
I_{CS}=-\frac{\kappa^2 h}{2 \pi \lambda^4} Q_4\wedge Q_4~.
\label{ICS}
\ee
However, these are not the only sources which contribute to the
anomaly. In particular, we expect a term (see Appendix B) 
\be
S_R=-\xi_R \int_{M^7} A_3\wedge {\rm tr}R^2~, \label{SR} 
\ee 
in the 7D action where $\xi_R$ is a dimensionful constant. As explained 
in Appendix B, such a term exists in the gauged 7D ${\cal N}=2$ supergravity 
theory resulting from the $K3$ compactification of the 11D five-brane 
anomaly term, and is expected to survive after gauging. The anomalous 
variation of the term (\ref{SR}) is
\be
\delta S_R=-\frac{\kappa^2\xi_R }{ \sqrt{2}
\lambda^2} \int_{M^6}Q_2^1\wedge {\rm tr}R^2~,
\ee
so that the corresponding anomaly eight-form becomes
\be
\label{IR}
I_R=-\frac{\kappa^2\xi_R }{2\pi \sqrt{2} \lambda^2}Q_4\wedge
{\rm tr}R^2~. 
\ee
In addition, as follows from Section 3.1, the appropriately normalized 
anomaly eight-form for the boundary theory is
\be
I_{bdy}=\frac{1}{(2\pi)^4}\left[\frac{1}{4608}\Big{(}N_V\!-\!N_H\!
+\!7N_T\Big{)}({\rm tr}R^2)^2\!-\!\frac{\beta}{96}{\rm tr}R^2{\rm
tr}F^2\!+\!\frac{\gamma}{24}({\rm tr}F^2)^2\right]~.
\ee
The reducible part of the anomaly eight-form from the bulk fields 
(\ref{anomdist}), is evenly distributed between the two fixed points 
and contributes a term
\be
I_{bulk}=\frac{1}{(2\pi)^4}\frac{-1}{2\cdot 5760}\frac{225}{4}({\rm tr}R^2)^2~,
\ee
at each fixed point.
Finally, there exists a  contribution to the anomaly arising 
from the gravitational Chern-Simons term (\ref{SGCS}). The irreducible 
${\rm tr}R^4$ part in Eq.(\ref{IGCS}) has been cancelled against the 
bulk and boundary irreducible parts of the anomaly. 
The remaining contribution of the gravitational Chern-Simons term to the 
anomaly is then
\be
I_{GCS}=\frac{1}{(2\pi)^4}\frac{n}{8\cdot 192}({\rm tr}R^2)^2~.
\label{IGSCR}
\ee
The total anomaly eight-form coming from the bulk, the boundary theory and the Chern-Simons terms is
\be
I_{{\rm total}}=I_{bulk}+I_{bdy}+I_{CS}+I_{R}+I_{GCS}~.
\ee
It is a polynomial in $({\rm tr}R^2)^2,~ {\rm tr}R^2{\rm
tr}F^2$ and $({\rm tr}F^2)^2$ and the vanishing of the total anomaly is
equivalent to the vanishing of the coefficients of these terms.
In particular,  the vanishing of the $({\rm tr}R^2)^2$ and $ {\rm tr}R^2{\rm
tr}F^2$ terms gives the conditions
\be
&&320\gamma\xi_{CS}^2-80\beta \xi_{CS}+2(N_V-N_H+N_T)+3=0\, , 
\label{c11}
\\
&& 384 \sqrt{2} \pi^3 \xi_R
\frac{\kappa^2}{\lambda^2}=\beta-8\gamma\xi_{CS}~, \label{c12}
\ee
respectively, whereas the vanishing of the $({\rm tr}F^2)^2$
specifies the dimensionless ratio, $\eta$,  as
\be
 \eta\equiv\frac{h\kappa^2}{\lambda^4}=\frac{\gamma}{3 (4\pi)^3}~.
\label{etarel}
\ee
This relation fixes the gauge coupling, $\lambda$,
in terms of the gravitational coupling, $\kappa$, and the topological
mass parameter, $h$, of the Chern-Simons term. This is similar to the
relation obtained in the HW theory except for the presence of the extra
parameter, $h$. The difference is due to the fact that in the 11D HW theory
the Chern-Simons term is fixed by supersymmetry, whereas in seven dimensions
the theory is supersymmetric up to an arbitrary topological mass
parameter, $h$. 

Note that similar conditions to (\ref{c11}),(\ref{c12}),
and (\ref{etarel}) exist at the second fixed point $x^7=\pi R$. 
When the boundary theories at the two fixed points are the same, 
the anomaly cancellation conditions at $x^7=\pi R$ are identical to that 
at $x^7=0$. However, if the boundary theory at $x^7=\pi R$ is different 
from the boundary theory at $x^7=0$, then the anomaly cancellation 
conditions must be solved for another set of $\xi_{CS},\lambda,\beta$, 
and $\gamma$. In this case a different value of $\xi_R$ is also needed 
at the second fixed point. This requires more general compactifications 
of the 11D theory.

The effective gauge coupling on the boundaries,
$\lambda_{\rm eff}^2\equiv\lambda^2 \langle\sigma^2\rangle$ can be 
written in terms of the geometric parameters as
\be
\lambda_{\rm eff}^2=2\kappa^2 \sqrt{\frac{6\pi^3\Lambda_{(0)}}{5\gamma}}~.
\ee
We see then that $\gamma$ must be necessarily positive in order
to have no ghosts. In this case, decoupling gravity leads to
an anomaly free theory as pointed out in~\cite{Seiberg:1996qx}.

\subsubsection{Boundary hypermultiplets}

In the previous section we only considered boundary vector
multiplets and their bulk couplings. We will now introduce
hypermultiplets on the boundary and determine their supersymmetric
interactions with bulk fields. Let us begin with the globally
supersymmetric action for the hypermultiplet
$(\varphi^\alpha,\zeta_Y)$ \be S_H^{(0)}=\int d^6x
\sqrt{-g}\left(-\frac{1}{ 2}\,
g_{\alpha\beta}(\varphi)\partial_\mu\varphi^\alpha\partial^\mu\varphi^\beta
-\frac{1}{ 2}\bar{\zeta}^Y\Gamma^\mu D_\mu\zeta_Y\right)~.
\label{hyperaction} \ee Here,
$\varphi^\alpha~(\alpha,...=1,...,4n_H)$,
$\zeta^Y~(Y,...=1,...,2n_H)$ are, respectively, the scalar and
fermion components of the $n_H$ hypermultiplets on the boundary,
and $g_{\alpha\beta}$ is the metric of the scalar manifold. The
action (\ref{hyperaction}) is the standard ${\cal N}=(0,1)$
globally supersymmetric action where the $4n_H$ scalars
parametrize a space with $Sp(n_H)$ holonomy group, i.e. a
hyperk\"ahler manifold. The covariant derivative in
(\ref{hyperaction}) is defined as \be
D_\mu\zeta^Y=\partial_\mu\zeta^Y+\Gamma^Y_{\alpha X}\partial_\mu
\varphi^\alpha \zeta^X~, \ee where $\Gamma^Y_{\alpha X}$ is the
$Sp(n_H)$ connection. We will demand invariance of the action
(\ref{hyperaction}) under the supersymmetry transformations \be
&&\delta\varphi^\alpha=\frac{1}{ 2} f_\varphi
{V^\alpha}_{iY}\bar{\epsilon}^i\zeta^Y~,\label{susyhyp1} \\ &&
\delta\zeta^Y=\frac{1}{ 2} f_\zeta\ {{V_{\alpha i}}^Y}  \Gamma^\mu
\partial_\mu \varphi^\alpha\epsilon^i~.
\label{susyhyp2} \ee where $f_\varphi,f_\zeta$ are also functions
of the bulk fields, and ${{V_{\alpha i}}^Y}$ is the vielbein of
the scalar manifold. The case $f_\varphi=f_\zeta=1$ corresponds to
the ${\cal N}=(0,1)$ globally supersymmetric theory. By simple
inspection of the transformations
(\ref{susyhyp1})-(\ref{susyhyp2}), we find that
$f_\varphi=f_\zeta^{-1}$ in order that two consecutive
supersymmetry transformations on $\varphi^\alpha$ produce a
correctly normalized translation. Checking the supersymmetry of
the action (\ref{hyperaction}), we find that ${V_{\alpha i}}^Y$ is
covariantly constant so that the vielbein satisfies the relations
\be
g_{\alpha\beta}{V^\alpha}_{iY}{V^\beta}_{jZ}&=&\epsilon_{ij}\epsilon_{YZ}~,\\
{V^\alpha}_{iY}{V^{\beta jY}}+{V^\beta}_{iY}V^{\alpha jY}
&=&g^{\alpha\beta}\delta_i^j~, \\
{V^\alpha}_{iY}{V^{\beta i Z}}+{V^\beta}_{iY}V^{\alpha iZ}
&=&\frac{1}{ n_H}g^{\alpha\beta}\delta_Z^Y~,
\ee
where $\epsilon_{ij}$ and $\epsilon_{YZ}$ are the $Sp(n_1)$ and
$Sp(n_H)$ invariant antisymmetric tensors, respectively.
We can now define a triplet of complex structures
${{J_{\alpha\beta}}_i}^j$ as
\be
{{J_{\alpha\beta}}_i}^j={V_\alpha i}^{Y}{{V_\beta}^j}_ Y-
{V_\beta i}^{Y}{{V_{\alpha}}^j}_ Y \label{JJ}~,
\ee
which obey the $SU(2)$ algebra, and are covariantly constant with respect
to the $Sp(1)$ connection ${\omega_i}^j={\omega_{\alpha i}}^jd\varphi^\alpha$,
namely
\be
\nabla {J_i}^{j}=d{J_i}^{j}+{\omega_i}^k
{J_k}^j-{J_i}^k{\omega_k}^j=0~.
\ee
The $Sp(1)$ curvature two-form
${\Omega_i}^j=\frac{1}{ 2}{\Omega_{\alpha\beta i}}^{j}
d\varphi^\alpha \wedge d\varphi^\beta$ is
\be
 {\Omega_i}^{j}= d{\omega_i}^j+{\omega_i}^k\wedge {\omega_k}^j~,
\ee
and the manifold is then quaternionic if
\be
{\Omega_i}^{j}=\mu \, {J_i}^j~,
\ee
for some constant $\mu\neq 0$, whereas it is hyperk\"ahler if
$\mu=0$, i.e. a hyperk\"ahler manifold has vanishing $Sp(1)$
curvature. So far we have seen that the holonomy group of the
scalar manifold should be contained in $Sp(n_H)\times SU(2)$ as
follows from the covariantly constant vielbein ${{V_{\alpha
i}}^Y}$. However, after imposing the latter condition, there
still exists an uncancelled term in the variation of
(\ref{hyperaction}) under the transformations
(\ref{susyhyp1})-(\ref{susyhyp2}). This term is explicitly written as
\be
\Delta^{(0)}_H=  \frac{1}{ 2}\int d^6x
\sqrt{-g}\ \bar{\epsilon}^i\Gamma^\mu\Gamma^\nu\partial_\mu\varphi^\alpha
 V_{\alpha i Y}\zeta^Y \partial_\nu f_\zeta~,
\label{d0}
\ee
and we will return to this term shortly.

As usual for local supersymmetry, we should add
to the action (\ref{hyperaction}) the standard $\bar{\psi}_\mu{\cal{S}}_H$
interaction where ${\cal{S}}_H$ is the supercurrent of the
hypermultiplet. In particular, the term which has to be added is
\be
S^{(1)}_H=\frac{1}{ 2} \int d^6x \sqrt{-g}\,f_\zeta \bar{\psi}_\mu^i\ \Gamma^\nu
\Gamma^\mu \partial_\nu \varphi^\alpha{V_{\alpha i}}^Y\zeta_Y~.
\ee
However, due to the variation of $\zeta^Y$ in
$S^{(1)}_H$, we get an uncancelled term which is
\be
\Delta_H^{(1)}=-\frac{1}{ 8}\int d^6 x \sqrt{-g}\,f_\zeta^2\bar{\psi}_\mu^i\
\Gamma^{\mu\nu\rho}\partial_\nu \varphi^\alpha\partial_\rho\varphi^\beta
\left({V_{\alpha i}}^YV_{\beta j Y}-{V_{\alpha j}}^YV_{\beta i Y}\right)
\epsilon^j~.
\ee
The only way to cancel $\Delta_H^{(1)}$ is to find an opposite
contribution
from the bulk. As in the vector case, there exists a
bulk term $\bar{\psi}_K^i\Gamma^{KMNR}\psi_{Rj}{F_{MN_i}}^j$, which
can be shown to be supersymmetric by invoking a Bianchi identity.
Using this term we may cancel $\Delta_H^{(1)}$ in the
downstairs approach by modifying the Bianchi identity to be
\be
{D F_{7\mu\nu i}}^j=\frac{i\kappa^2}{2 \sqrt{2}} \delta(x_7)
\partial_\mu \varphi^\alpha\partial_\nu\varphi^\beta
{J_{\alpha\beta i}}^j~,
\label{bb}
\ee
with $f_\zeta=\sigma^{1/2}$. The latter is also  needed for cancelling
(\ref{d0}) as we will see below. There are also other terms
that need to be cancelled in $S^{(1)}_H$, which arise from the
variation of $\bar{\psi}_\mu$, and are of the form
$F_{\mu\nu\rho7}\partial\varphi\epsilon$ and
${F_{\mu7 i}}^{j}\partial\varphi\epsilon$. The former can be cancelled
by adding
\be
&&S_H^{(2)}=\frac{1}{ 2\sqrt{5}}\int d^6x \sqrt{-g}\ \sigma^{1/2}
V_{\alpha i Y}\bar{\zeta}^Y\Gamma^\mu\partial_\mu \varphi^\alpha
\chi^i~, \\&&
S^{(3)}_H=\frac{1}{  24\sqrt{2}}\int d^6x \sqrt{-g}\ \sigma^{1/2}
\bar{\zeta}^Y\Gamma^{\mu\nu\rho}\zeta_YF_{7\mu\nu\rho}~.
\ee
In addition, the variation of $S^{(2)}$ with $\delta \chi\sim
\partial \varphi \epsilon$ exactly cancels (\ref{d0}) for
$f_\zeta=\sigma^{1/2}$ as promised.

Finally there remains the cancellation of terms of the form
$\zeta{F_{\mu7 i}}^{j}\partial\varphi\epsilon$ which arise from
variations of $\psi_\mu$ and $\chi$ proportional to
${F_{\mu7 i}}^{j}\epsilon$ in $S_H^{(1)}$ and $S_H^{(2)}$, respectively.
The uncancelled terms are of the form ${V_{\alpha i}}^Y\epsilon^j
\Gamma^{\mu\nu}\zeta_Y \partial_\mu\varphi^\alpha {F_{\nu 7j}}^i$ and
${V_{\alpha i}}^Y\epsilon^j \zeta_Y\partial^\mu\varphi^\alpha {F_{\mu
7j}}^i$. The former cancels exactly while the latter does not and so
we have to look for a possible bulk contribution. We need to implement
a correction to the supersymmetry variation of
${F_{\mu 7j}}^i$ which then produces a contribution from the kinetic term of
the two-form, just as there was in the vector multiplet case
from the kinetic term of the four-form. The correction to the supersymmetry
variation of ${F_{\mu 7j}}^i$ is
\be
\tilde{\delta}{F_{\mu7j}}^i= \frac{\kappa^2}{ 5 \sqrt{2}}\sigma^{-1/2}
\partial_\mu \varphi^\alpha\left({V_{\alpha i}}^Y\epsilon^j-
\frac{1}{ 2}\delta^j_i {V_{\alpha k}}^Y\epsilon^k\right)\zeta_Y~.
\ee
Summarizing, the supersymmetric boundary action for neutral
hypermultiplets is
\be
 S_H&=&\int d^6x \sqrt{-g}\left[-\frac{1}{ 2}
g_{\alpha\beta}(\varphi)\partial_\mu\varphi^\alpha\partial^\mu\varphi^\beta
-\frac{1}{ 2}\bar{\zeta}^Y\Gamma^\mu D_\mu\zeta_Y \right.\nonumber\\
&&+\left.
\frac{1}{ 2}\sigma^{1/2} \bar{\psi}_\mu^i\ \Gamma^\nu
\Gamma^\mu \partial_\nu \varphi^\alpha{V_{\alpha i}}^Y\zeta_Y+
\frac{1}{ 2\sqrt{5}} \sigma^{1/2}
V_{\alpha i Y}\bar{\zeta}^Y\Gamma^\mu\partial_\mu \varphi^\alpha
\chi^i \right.\nonumber \\
&&+\left.\frac{1}{  24\sqrt{2}} \sigma^{1/2}
\bar{\zeta}^Y\Gamma^{\mu\nu\rho}\zeta_YF_{7\mu\nu\rho}\right]~,
\label{hyperaction1}
\ee
and it is invariant under the superymmetry transformations
\be
&&\delta\varphi^\alpha=\frac{1}{ 2} \sigma^{-1/2}
{V^\alpha}_{iY}\bar{\epsilon}^i\zeta^Y~,\\ &&
\delta\zeta^Y=\frac{1}{ 2}  \sigma^{1/2}\ {{V_{\alpha i}}^Y}  \Gamma^\mu
\partial_\mu \varphi^\alpha\epsilon^i~.
\ee

Let us now return to the scalar manifold where  the only
constraint  we have so far is that the holonomy group is in
$Sp(n_H)\times Sp(1)$. In 6D (as well as ${\cal N}=2$ in 4D),
${\cal N}=(0,1)$ supersymmetry requires that the scalar manifold
be hyperk\"ahler in rigid supersymmetry and quaternionic for local
supersymmetry. It is interesting to see how the quaternionic
structure arises. In the local case, the supersymmetry parameters
are charged (with minimal coupling) under $SU(2)$, and after a
supersymmetry transformation a term proportional to the $SU(2)$
curvature is generated from the gravitino kinetic term leading to
a quaternionic scalar manifold. In our case, the gravitino is not
minimally coupled to  the $SU(2)$ connection of the boundary
scalar manifold, and it is not clear how to obtain a quaternionic
structure as required by ${\cal N}=(0,1)$ 6D supersymmetry.

Although the 7D gravitino kinetic term does not contribute to the
$Sp(1)$ curvature of the scalar manifold, there is a contribution from
the Bianchi identity (\ref{bb}). Indeed, either from the Bianchi
identity or in the downstairs approach, we find that the boundary value of
the $SU(2)$ bulk gauge field is
\be
 {F_{\mu\nu i}}^j=\frac{i}{ 4 \sqrt{2}}\kappa^2\partial_\mu \varphi^\alpha
 \partial_\nu\varphi^\beta {J_{\alpha\beta i}}^j~.
\ee
It is more convenient to use form notation so that the boundary value
of the $SU(2)$ field strength can be rewritten in the form
\be
{F_i}^j|_{\partial M^7}=\frac{i}{ 4 \sqrt{2}}\kappa^2 {(\varphi^*J)_i}^j~,
\label{fij}
\ee
where, as usual, $(\varphi^*J)$ is the pullback of $J$ on $M^7$ defined
as
\be
   {(\varphi^*J)_i}^j=\frac{1}{2}{J_{\alpha\beta i}}^j\,  \partial_\mu
     \varphi^\alpha\partial_\nu\varphi^\beta\,dx^\mu\wedge dx^\nu~,
\ee and similarly for higher-order forms. Since we have that
$D{F_i}^j=\nabla {J_i}^j=0$, we obtain \be {A_i}^j|_{\partial
M^7}=-\frac{i}{ g}{(\varphi^*\omega)_i}^j~, \ee where
${(\varphi^*\omega)_i}^j$ is the pullback of the $Sp(1)$
connection in the scalar manifold. As a result, the boundary value
${F_i}^j=d{A_i}^j+i g{A_i}^k\wedge {A_k}^j $ is proportional to
the pullback of the $Sp(1)$ curvature two-form of the  scalar
manifold \be {F_i}^j|_{\partial M^7}=-\frac{i}{
g}{(\varphi^*\Omega)_i}^j~. \ee Then, using Eq.(\ref{fij}) we
obtain \be {\Omega_i}^j=- \frac{g \kappa^2}{ 4\sqrt{2}}{J_i}^j~,
\ee so that, the scalar manifold is indeed quaternionic as
required by ${\cal N}=(0,1)$ 6D local supersymmetry.

\section{Phases of the Boundary Theory }

We have seen that the supersymmetric vacua of the ${\cal N}=2$ 7D
theory are pure $AdS_7$, and the two-brane Randall-Sundrum
configuration (RS1). There also exists the one-brane
Randall-Sundrum vacuum (RS2)~\cite{rs}, where a single six-brane
is sitting at $x_7=0$. It is obtained by omitting the last
$\epsilon$ and $\delta$ functions in eqs.(\ref{h}) and (\ref{Ss}),
respectively. The metric is still given by (\ref{rs1}), except
that now $0\leq y<\infty$, and the singularity at $y=0$ is
resolved by placing a positive-tension brane there. In this case,
only the graviton is localized on the brane so that there exists
an ${\cal N}=(0,1)$ 6D theory on the brane. Again, the 6D theory
is anomalous (since Eq.(\ref{irredcond}) is violated) and we
should again include matter fields on the brane to cancel the
anomaly by a local GS mechanism at the single point $y=0$ as we
did before. The only differerence with the two-brane RS1 vacuum is
that here we have only local cancellation of the anomaly since
global and local cancellations coincide (only one brane). Thus,
the irreducible part of the anomaly cancels if
Eq.(\ref{irredcond}) is satisfied and then the rest of the anomaly
should be cancelled by a 6D GS mechanism. As a result, any anomaly
free 6D theory can be put on the brane leading to a consistent
$\R/\Z_2$ ``compactification'' of the 7D ${\cal N}=2$ theory.
There are many 6D theories which can exist on the boundary. In the
case of one tensor multiplet, some of them can be obtained from
heterotic string compactifications like $E_8\times E_7$, and
others like $SU(n)\times SU(n)$ with matter in the representation
$({\bf n},{\bf \bar{n}})+({\bf \bar{n}},{\bf n})+242 (1,1)$
described in \cite{Schwarz:1995zw} do not arise from any known
compactifications of string theory. In the case of more than one
tensor multiplets, a generalized GS mechanism may operate as in
\cite{Sagnotti:1992qw}. As a result, any anomaly free ${\cal
N}=(0,1)$ 6D theory can be lifted to the boundary of a 7D ${\cal
N}=2$ one-brane vacuum. The boundary 6D theory can be consistently
coupled to the 7D bulk in a similar fashion as considered in the
previous section.

In the two-brane scenario, it is also interesting to consider the
two possible limits $R\to 0$, $R \to \infty$ in the  $n_T=1$ case. These
limits correspond to coincident branes and a single brane,  respectively.
As we have discussed earlier, the anomaly must be factored both globally
and locally. In particular, we have seen that, due to Eq.(\ref{sss})
the local anomaly (\ref{gs}) can be written in the form
\be
\label{ptfac}
    I_8^{(i)}=\left(c_i{\rm tr}R^2+|a_i|
      {\rm tr}F_i^2\right) \left({\rm tr}R^2-|b_i|
      {\rm tr}F_i^2\right)~,
\ee
where $|a_i||b_i|=\frac{2}{3}\gamma_i>0$.
Then it is possible that a phase transition on the boundary may
occur. This is because the factorization (\ref{ptfac}) is
correlated with the gauge kinetic terms on the boundary
via~\cite{Sagnotti:1992qw}
\be
   -\left(|a_i| \sigma^{-2} -|b_i| \sigma^2\right) {\rm tr} F_i^2~,
\ee
and the gauge coupling on the boundaries always becomes infinite at
the value
\be
     \sigma^4 = \frac{|a_i|}{|b_i|}~.
\ee
As the radius of the $S^1/\Z_2$ compactification is
$R\sim \sigma^{-2}$, there is always a value of $R$ where the gauge
coupling blows up and a phase transition takes place.
Thus, one of the limits $R\to 0$ or $R~\to~\infty$ drives the gauge
theory  to   a phase transition where the number of massless
degrees of freedom rearrange themselves such that the 6D theory remains
anomaly free. One of the possibilities for this singularity is that
it could signal the appearance of tensionless
strings~\cite{Seiberg:1996vs,Duff:1996cf}.
This behaviour is novel and it does not occur in the HW scenario.

Although the coincident limit of the two branes is unique,
the limit $R\to \infty$ can be taken in two ways.
We can either send the UV or the IR brane to infinity.
When the IR brane is sent to infinity the localized theory
on the UV-brane is simply the single brane vacuum
considered above, and the 6D boundary theory is anomaly free.

On the other hand when we send the UV brane to infinity there is
no longer any localized graviton on the IR brane. This is due to
the fact that the bulk theory is $AdS_7$. Thus there are no
gravitational anomalies in the 6D boundary theory. However there
are still vector and tensor multiplets on the boundary which
give rise to gauge anomalies. Assuming that the hypermultiplet and
tensor also decouple (since they arise from the 7D gravity multiplet),
we are left with $n_V$ vector multiplets, one tensor multiplet and
one neutral hypermultiplet on the IR brane where the latter are
localized in any case there. The theory is then anomaly free as
long as $\gamma_2>0$ as discussed in~\cite{Seiberg:1996qx}.

There is also a dual description of our anomaly free 7D
brane worlds. This is similar to the HW model where the dual
description of the 11D supergravity is the ten-dimensional strongly
coupled $E_8\times E_8$ heterotic string theory. Since the vacua of the
7D brane worlds is $AdS_7$, we simply rely on AdS/CFT
correspondence~\cite{malda,ver,Gub,adsbound} to identify the dual description
as a 6D strongly coupled CFT. These field theories are very interesting since
by the AdS/CFT dictionary the fields on the UV brane correspond to
fundamental fields added to the CFT, while fields on the IR brane are
bound states of the CFT. For example, since  the gauge group is in general
a product group ${\cal G}_1\times {\cal G}_2$, this means that the
${\cal G}_1$ gauge fields are fundamental, while the ${\cal G}_2$ gauge fields
are composite in the dual description. In addition since gravity
is localized on the UV brane, we must add a 4D gravity multiplet to the
dual field theory. Unlike the HW case, where the dual theory is identified
as the 10D heterotic string theory, not much is known about the dual
6D theories of our 7D brane worlds. These await further investigation.

\section{Conclusions}

We have constructed anomaly-free seven-dimensional brane worlds
based on the minimal ${\cal N}=2$ 7D gauged supergravity. Under an
$S^1/\Z_2$ compactification, we showed that the resulting spectrum
is the chiral ${\cal N}=(0,1)$ 6D supergravity. In order to
maintain supersymmetry, we introduced two six-branes of opposite
tension at the orbifold fixed points. Since there exists a
negative bulk cosmological constant, this leads to a 7D vacuum
solution which is a slice of $AdS_7$. However, unlike the
celebrated 5D Randall-Sundrum solution, where there are no
constraints on the boundary theory, in our case, anomaly
cancellation puts stringent restrictions on the possible boundary
matter content. This is because there are gravitational anomalies
in six dimensions. This is analogous to the HW model, where the
cancellation of 10D gravitational anomalies uniquely specifies the
boundary theory to be $E_8\times E_8$ with an $E_8$ factor
localized on each boundary. In contrast, while there are many
anomaly-free 6D theories, the boundary matter content of our 7D
brane world is nevertheless constrained. In particular, for the
case of one tensor multiplet and gauge groups containing the
standard model, only exceptional groups are allowed on the
boundaries with a further restriction on the number of fundamental
generations. If one allows $n_T>1$ then using the Sagnotti
mechanism many more possibilities exist for the boundary theory.

We also explicitly constructed the locally supersymmetric bulk-boundary
Lagrangian, up to fermionic bilinear terms, for a boundary vector
multiplet and hypermultiplet. The case of the vector multiplet
relies on the modification of the Bianchi identity of the four-form
field strength. This is very similar to what happens in
the HW model. Moreover, anomaly cancellation fixes a dimensionless ratio
$\eta$, formed from the boundary gauge coupling, the 7D gravitational
coupling, and the topological mass parameter of the Chern-Simons term. Again
this is analogous to a similar relation in the HW theory, except that now
there is a dependence on the extra parameter from the Chern-Simons term.
However, the case of boundary hypermultiplets has no counterpart in the
11D HW model. For these boundary fields we had to modify the Bianchi
identity of the bulk gauge field resulting from the gauged R-symmetry.
This modification was also crucial in showing that the scalar manifold
of the boundary hypermultiplets becomes quaternionic, as expected
in the locally supersymmetric limit.

The brane worlds constructed in this paper are a first step in obtaining
boundary theories which have the standard model matter content. For example,
one can envisage compactifying the six-branes in our anomaly-free models
on the sphere $S^2$ with a monopole background~\cite{mono} to obtain
an effective four-dimensional chiral theory. This would be the counterpart
of the Calabi-Yau compactifications of M-theory, except that in our case the
bulk space is AdS, and not Minkowski. In addition there is also the
intriguing question of understanding the
dual formulation of our models. By the AdS/CFT correspondence our 7D
brane worlds are dual to a class of 6D strongly coupled conformal field
theories. These 6D theories must necessarily include fundamental fields
associated with the localized fields on the UV brane, such as the 4D gravity
multiplet, and any gauge and matter fields, while for the fields
localized on the IR brane, they will appear as bound states of the CFT.
Our 7D AdS brane worlds suggest a way to study these mysterious hybrid
6D theories further.

\section*{Acknowledgements}
We thank C. Angelantonj, E. Kiritsis, J. March-Russell, E. Poppitz,
R. Rattazzi, A Sagnotti, J. Sonnenschein, and R. Sundrum
for discussions. The work of TG was supported in part by a DOE grant
DE-FG02-94ER40823 at the University of Minnesota. The work of AK was
supported in part by the European RTN networks HPRN-CT-2000-00122 and
HPRN-CT-2000-00131, and the Greek State Foundation
Award ``Quantum Fields and Strings" (IKYDA-2001-22).
TG wishes to acknowledge the Aspen Center for Physics
where part of this work was done.

\section*{Appendix A}

The $8\times 8$ 6D gamma matrices satisfy the Clifford algebra
\be
\{\gamma^\alpha,\gamma^\beta\}=2\eta^{\alpha\beta}\, , ~~~
\alpha,\beta=0,1,...,5
\ee
with $\eta_{\alpha\beta}=\mbox{diag}(-,+,...,+)$. The matrix $\gamma^7$ is
defined as
\be
\gamma^7=\gamma^0\gamma^1...\gamma^5\, , ~~~~~~({\gamma^7})^2=1~,
\ee
and the $8\times 8$  matrices $\Gamma^A=(\gamma^\alpha,\gamma^7)$
satisfy the 7D Clifford algebra
\be
\{\Gamma^A,\Gamma^B\}=2 \eta^{AB}\, , ~~~ A,B=0,1,...6
\ee
All 7D spinors are symplectic-Majorana
\be
\chi^i=\epsilon^{ij}{\bar{\chi}_j}^T\, ,
~~~\bar{\chi}_i={\chi^i}^\dagger\Gamma_0\, ,
\ee
and $SU(2)$ indices $i,j=1,2$ are raised and lowered as
\be
\chi^i=\epsilon^{ij}\chi_j\, , ~~~\chi_i=\chi^j\epsilon_{ji}\, ,
~~\epsilon^{ij}=\epsilon_{ij}=\left(\begin{array}{cc}
0&1\\
-1&0
\end{array} \right)~.
\ee
A 7D spinor $\chi^i$ decomposes into $\chi^i=\chi^i_++\chi^i_-$,
where $\chi^i_\pm$ are 6D symplectic Majorana-Weyl spinors
satisfying $\gamma_7\chi^i_\pm=\pm \chi^i_\pm$.
Contraction with the $SU(2)$-invariant  antisymmetric tensor
is always understood in spinor inner-products, e.g.
$\bar{\chi}\Gamma^{ABC}\psi=\bar{\chi}^i\Gamma^{ABC}\psi_i$. As a
result we have
\be
\bar{\chi}\Gamma^{A_1...A_n}\psi=(-1)^n\bar{\psi}\Gamma^{A_n...A_1}\chi~.
\ee
The same conventions also hold for 6D spinors. Finally, the
Riemann tensor is ${R^{AB}}_{MN}=\partial_M
{\omega_{N}}^{AB}+{\omega_M}^{AC}{\omega_{NC}}^B-{ {M\leftrightarrow
N}\mathbb{}}$, and the Ricci scalar $R={R^{AB}}_{MN}{e_A}^M{e_B}^N$.

\newpage
\section*{Appendix B}

Let us now see how the dimensional reduction of
the interaction term $F_4\wedge X_7$ of 11D supergravity gives rise to
a gravitational Chern-Simons term, $S_{GCS}$ and the term $S_R$ in (\ref{SR}).
In 11D supergravity  the correct normalization of
the $F_4\wedge X_7$ term is
\be
S_{GCS}=\frac{1}{2}\left(\frac{4\pi^2}{3 \kappa_{11}^2}\right)^{1/3}
\int F_4\wedge X_7~,
\ee
where $\kappa_{11}$ is the 11D Newton's constant.
Upon compactification on $K3$, the above interaction gives rise to
two terms in the 7D effective supergravity action
\be
S_{GCS}=\frac{1}{2}\left(\frac{4\pi^2}{3 \kappa_{11}^2}\right)^{1/3}\left(
\int_{K3} F_4\right)\int X_7\equiv \xi_G \int X_7~,
\ee
and
\be
\label{11dSR}
S_{R}=\frac{1}{2}\left(\frac{4\pi^2}{3 \kappa_{11}^2}\right)^{1/3}
\int F_4 \wedge \int_{K3} X_7~.
\ee
According to \cite{Witten:1996md},\cite{Bilal:1999ig}, the $F_4$-fluxes 
are quantized as
\be
\int_{{\cal{C}}_4}F_4=(6 \pi)^{1/3}\kappa_{11}^{2/3}\, n~,
\ee
where $n$ is an integer or half-integer.
If ${\cal{C}}_4$ is the $K3$ surface, we obtain for $S_{GCS}$ the result
\be
S_{GCS} = n \pi \int X_7~, \label{SGCS}
\ee
so that $\xi_G=n \pi$ and $X_7$ is a 7D Chern-Simons term satisfying
\be
X_8=dX_7=\frac{1}{(2\pi)^4}\frac{1}{192}\left[\frac{1}{4}({\rm tr}R^2)^2
-{\rm tr}R^4\right]~.
\ee
Under diffeomorphisms, $X_7$ transforms as $X_7\to X_7+dX_6^{(0)}$.
Thus, on a manifold with a boundary, the anomalous variation of $S_{GCS}$
is
\be
\delta S_{GCS}=\xi_G\int \, X_6^{(0)}\, ,
\ee
and the corresponding anomaly eight form is then
\be
I_{GCS}=\frac{\xi_G}{2\pi} \, X_8\left[\delta(y)-\delta(y-\pi R)\right]~, 
\label{IGCS}
\ee
and similarly at $y=\pi R$.

Performing the integral in (\ref{11dSR}), it is not difficult to see that
\be
X_3=\int_{K3}X_7= \frac{1}{4\cdot (2\pi)^2} \omega_{3L}~,
\ee
so that the term $S_R$ can be written as
\be
S_R=-\frac{1}{8\cdot (2\pi)^2}\left(\frac{4\pi^2}{3 \kappa_{11}^2}\right)^{1/3}
\int A_3 \wedge {\rm tr}R^2~.
\ee
Thus the coefficient $\xi_R$ in (\ref{SR}) is determined to be
\be
\xi_R=\left(\frac{\pi^2}{48\kappa_{11}^2}\right)^{1/3}=\left(\frac{\pi^2}
{48\kappa^2}\right)^{1/3}{V_{K3}}^{-1/3}~,
\ee
where $V_{K3}$ is the volume of $K3$ and the relation
$\kappa_{11}^2=V_{K3}\kappa^2$ between the 11D and 7D Newton constants
$\kappa_{11}$ and $\kappa$, respectively, has been used.

\newpage
\section*{Appendix C}

We tabulate here all the solutions which satisfy the anomaly constraint
conditions for $n_T=1$, and matter content (\ref{hyprep}).

\begin{table}[!h]\centering
\begin{tabular}{|c|c|}\hline
${\cal G}_1\times {\cal G}_2$ & $(n_1,n_2,n_S)$\\ \hline\hline
$E_8\times E_7$ & (0,10,64), (1,5,96) \\ \hline
$E_8\times E_6$ & (0,18,83) , (1,8,105) \\ \hline
$E_8\times F_4$ & (0,17,101), (1,7,113) \\ \hline
$E_8\times G_2$ & (0,11,428) , (0,46,183), (1,16,145), (2,1,2) \\
\hline\hline
$E_7\times E_7$ & ($n_1,8-n_1, 61$)\\ \hline
$E_7\times E_6$ & ($n_1,14-2n_1, 76- 2n_1$), (2,7,153)\\ \hline
$E_7\times F_4$ & ($n_1,13-2n_1, 90-4n_1$), (2,6,160)\\ \hline
$E_7\times G_2$ & ($n_1,34-6 n_1, 152-14 n_1$), (1,12,250),(2,13,187),
(6,7,5)\\ \hline\hline
$E_6\times E_6$ & ($n_1,12-n_1,75$), (2,7,156)\\ \hline
$E_6\times F_4$ & ($n_1,11-n_1,87-n_1$), (2,6,163),(5,9,4), (7,1,158)\\
\hline
$E_6\times G_2$ & ($n_1,28-3n_1,139-6 n_1)$, (0,12,251), (2,13,190),\\
& (3,14,156), (5,22,46), (9,6,50), (10,7,16) \\ \hline\hline
$F_4\times F_4$ & $(n_1,10-n_1, 87)$, (1,6,165),(4,9,9)\\ \hline
$F_4\times G_2$ & $(n_1,25-3 n_1,134-5n_1)$, (1,13,192), (2,14,159),
\\& (4,22,51), (8,6,59), (9,7,26) \\ \hline\hline
$G_2\times G_2$ & ($n_1,20-n_1,131$), (1,14,166), (6,19,96),\\ &
(7,22,68), (8,28,19) \\ \hline\hline
\end{tabular}
\end{table}

\end{document}